\documentclass[longauth]{aa}  
\usepackage{graphicx}
\usepackage{txfonts}
\usepackage{natbib}
\usepackage{multirow}
\newcommand{\pycs}{{\tt PyCS}\xspace}

\newcommand{\hc}{$H_0$}

\newcommand{\HEzerozero}{HE\ 0047$-$1756\xspace}
\newcommand{\WGzero}{WG\ 0214$-$2105\xspace}
\newcommand{\DESzero}{DES\ 0407$-$5006\xspace}
\newcommand{\Jzerohuit}{SDSS\ J0832$+$0404\xspace}
\newcommand{\deuxM}{2M\ 1134$-$2103\xspace}
\newcommand{\PSseize}{PSJ\ 1606$-$2333\xspace}
\newcommand{\DESvingt}{DES\ 2325$-$5229\xspace}
\newcommand{\DESvingttrentehuit}{DES\ 2038$-$4008\xspace}
\newcommand{\DESzeroquatre}{DES\ 0408$-$5354\xspace}
\newcommand{\PGonze}{PG\ 1115$+$080\xspace}
\newcommand{\WFIvingt}{WFI\ 2033$-$4723\xspace}

\begin{document}

\title{TDCOSMO II: Six new time delays in lensed quasars from high-cadence monitoring at the MPIA 2.2 m telescope\thanks{All light curves presented in this paper are only available in electronic form
at the CDS via anonymous ftp to \url{cdsarc.u-strasbg.fr} (130.79.128.5)
or via \url{http://cdsarc.u-strasbg.fr/viz-bin/cat/J/A+A/}}.}

\author{
%group 1
M.~Millon\inst{\ref{epfl}} \and
F. Courbin\inst{\ref{epfl}} \and
V.~Bonvin \inst{\ref{epfl}} \and
%Founders 
E. Buckley-Geer \inst{\ref{fermilab}, \ref{chicago}} \and
C.~D. Fassnacht \inst{\ref{ucdavis}} \and
J. Frieman \inst{\ref{fermilab}, \ref{chicago}} \and
P.~J. Marshall \inst{\ref{stanford}} \and
S.~H. Suyu \inst{\ref{MPG}, \ref{TUM}, \ref{ASIAA}} \and
T. Treu \inst{\ref{ucla}} \and
T. Anguita \inst{\ref{UNAB}, \ref{millenium}} \and
V. Motta \inst{\ref{valpo}} \and
A. Agnello \inst{\ref{DARK}} \and
%Observers
J.~H.~H.~Chan \inst{\ref{epfl}} \and
D.~C.-Y Chao \inst{\ref{MPG}, \ref{TUM}} \and
M. Chijani \inst{\ref{UNAB}} \and
D. Gilman \inst{\ref{ucla}} \and 
K. Gilmore \inst{\ref{stanford}} \and
C. Lemon \inst{\ref{epfl}} \and
J.~R. Lucey \inst{\ref{Durham}} \and
A. Melo \inst{\ref{valpo}} \and
E. Paic \inst{\ref{epfl}} \and
K. Rojas \inst{\ref{epfl}} \and
D.~Sluse \inst{\ref{Liege}} \and
P.~R. Williams \inst{\ref{ucla}} \and
A. Hempel \inst{\ref{UNAB}, \ref{heidelberg}} \and 
S. Kim \inst{\ref{puc}, \ref{heidelberg}}\and
R. Lachaume \inst{\ref{puc}, \ref{heidelberg}}\and 
M. Rabus \inst{\ref{LCOGT}, \ref{ucsantabarbara}}
}

\institute{
Institute of Physics, Laboratory of Astrophysics, Ecole Polytechnique 
F\'ed\'erale de Lausanne (EPFL), Observatoire de Sauverny, 1290 Versoix, 
Switzerland \label{epfl}\goodbreak \and
Fermi National Accelerator Laboratory, P.O. Box 500, Batavia, IL 60510, 
USA \label{fermilab}\goodbreak \and
Department of Physics, University of California, Davis, CA 95616, USA 
\label{ucdavis}\goodbreak \and
Kavli Institute for Cosmological Physics, University of Chicago, 
Chicago, IL 60637, USA \label{chicago}\goodbreak \and
Kavli Institute for Particle Astrophysics and Cosmology, Stanford 
University, 452 Lomita Mall, Stanford, CA 94035, USA 
\label{stanford}\goodbreak \and
Max Planck Institute for Astrophysics, Karl-Schwarzschild-Strasse
1, D-85740 Garching, Germany \label{MPG}\goodbreak \and
Physik-Department, Technische Universit\"at M\"unchen, 
James-Franck-Stra\ss{}e~1, 85748 Garching, Germany \label{TUM}\goodbreak 
\and
Institute of Astronomy and Astrophysics, Academia Sinica, P.O.~Box 
23-141, Taipei 10617, Taiwan \label{ASIAA}\goodbreak \and 
Department of Physics and Astronomy, University of California, Los 
Angeles, CA 90095, USA \label{ucla}\goodbreak \and
Departamento de Ciencias F\'isicas, Universidad Andres Bello Fernandez 
Concha 700, Las Condes, Santiago, Chile    \label{UNAB}\goodbreak \and
Millennium Institute of Astrophysics, Chile \label{millenium}\goodbreak  \and
Instituto de F\'isica y Astronom\'ia, Universidad de Valpara\'iso, Avda. 
Gran Breta\~na 1111, Playa Ancha, Valpara\'iso 2360102, Chile 
\label{valpo}\goodbreak \and
%Argelander-Institut f\"ur Astronomie, Auf dem H\"ugel 71, 53121, Bonn, Germany \label{bonn}\goodbreak \and
DARK, Niels Bohr Institute, University of Copenhagen, Lyngbyvej 2, 2100, Copenhagen, Denmark \label{DARK}\goodbreak \and
Centre for Extragalactic Astronomy, Department of Physics, Durham University, Durham DH1 3LE, UK \label{Durham} \goodbreak \and
%MIT Kavli Institute, Cambridge, MA 02139, USA \label{MIT} \goodbreak \and
Centro de Astroingenier\'ia, Facultad de F\'isica, Pontificia Universidad 
Cat\'olica de Chile, Av. Vicu\~na Mackenna 4860, Macul 7820436, 
Santiago, Chile \label{puc}\goodbreak \and
Max-Planck-Institut f\"ur Astronomie, K\"onigstuhl 17, 69117 Heidelberg, Germany \label{heidelberg}\goodbreak \and
STAR Institute, Quartier Agora - All\'ee du six Ao\^ut, 19c B-4000 Li\`ege, Belgium \label{Liege} \goodbreak \and
Las Cumbres Observatory Global Telescope, 6740 Cortona Dr., Suite 102, Goleta, CA 93111, USA \label{LCOGT} \goodbreak \and
Department of Physics, University of California, Santa Barbara, CA 93106-9530, USA \label{ucsantabarbara}
}

\date{\today}

\abstract{
We present six new time-delay measurements obtained from $R_c$-band monitoring data acquired at the Max Planck Institute for Astrophysics (MPIA) 2.2 m telescope at La Silla observatory between October 2016 and February 2020. The lensed quasars \HEzerozero, \WGzero, \DESzero, \deuxM, \PSseize, and \DESvingt were observed almost daily at high signal-to-noise ratio to obtain high-quality light curves where we can record fast and small-amplitude variations of the quasars. We measured time delays between all pairs of multiple images with only one or two seasons of monitoring with the exception of the time delays relative to image D of \PSseize. The most precise estimate was obtained for the delay between image A and image B of \DESzero, where $\tau_{AB} = -128.4^{+3.5}_{-3.8}$ d (2.8\% precision) including systematics due to extrinsic variability in the light curves.
For \HEzerozero, we combined our high-cadence data with measurements from decade-long light curves from previous COSMOGRAIL campaigns, and reach a precision of 0.9 d on the final measurement. The present work demonstrates the feasibility of measuring time delays in lensed quasars in only one or two seasons, provided high signal-to-noise ratio data are obtained at a cadence close to daily. 
}

\titlerunning{Six new time delays from high-cadence monitoring.}
\maketitle

%====================
\section{Introduction}
%====================
%
Time-delay cosmography with strongly lensed quasars was first proposed by \cite{Refsdal1964} as a single-step method to measure the Hubble constant \hc. The method relies on three ingredients. First, a precise measurement of the time delays between the lensed images must be obtained. This is typically achieved from photometric monitoring campaigns producing the light curve for each multiple image. Second, a mass model is needed for the main lensing galaxy and its possible companions. Deep and  high-resolution images, typically obtained with adaptive optics (AO) or the Hubble Space Telescope (HST) are needed for this task. Finally, we need to estimate the contribution of all intervening galaxies along the line of sight to the quasar. This last step can be performed statistically with galaxy counts in wide-field images \citep{Rusu2017}, direct multiplane modeling \citep{McCuly2017}, or weak lensing measurements \citep[e.g.,][]{Tihhonova2018}. These three ingredients allow for direct measurements of distances to the lens system, which together with the lens and source redshift measurements, provide constraints on $H_0$.

The method is complementary to other probes such as the cosmological microwave background (CMB), baryon acoustic oscillation (BAO), and the cosmic distance ladder, since time-delay cosmography is mainly sensitive to \hc~ and depends weakly on the other cosmological parameters. It is therefore an ideal probe to lift degeneracies in other experiments. Using lensed quasars, \cite{Wong2019} obtained a 2.4\% precision on the Hubble constant in flat-$\Lambda$CDM cosmology with a sample of six systems studied by the H0LiCOW collaboration \citep{Suyu2010, Suyu2014,Wong2017,Bonvin2017,Birrer2019,Rusu2019,Chen2019}. Combining this measurement with the latest results from the Cepheid distance ladder \citep{Riess2019}, the tension with the Planck results \citep{Planck2018} reaches 5.3$\sigma$, suggesting the presence of unaccounted systematics in one or both experiments or new physics beyond the $\Lambda$CDM model \citep[e.g.,][]{Verde2019, Riess2019b, Freedman2020}. 

%CSOMOGRAIL Programm and previous result
The COSMOGRAIL program has so far been one of the leading projects dedicated to time-delay measurement in strong lensing systems. This program produced decade-long light curves of more than 20 objects with 1 m class telescopes, yielding many precise time-delay measurements \citep[e.g.,][]{Tewes2013b, Eulaers2013, Rathna2013, Bonvin2017}. In particular, the final paper of the COSMOGRAIL series presents time delays for 18 objects \citep{Millon2020}. The observation strategy was recently enhanced with higher cadence (daily observation) and improved photometric precision and now allows us to catch quasar variations that are faster than the typical microlensing signal. Consequently, time delays can be measured to a few percent precision in only one monitoring season, provided 2 m-class telescopes can be used on a daily basis. This is the case of the MPIA 2.2 m telescope at ESO La Silla Observatory, which we use in the present work. Previous results using this telescope and strategy were presented in \cite{Courbin2017} and \cite{Bonvin2018a, Bonvin2019}. 

 In this paper, we report six new time delays with precisions in the range 2.8\% $< \delta(\Delta t) / \Delta t <$ 18.3\%. We first present in Sect.~\ref{sec:data} the high-cadence, high signal-to-noise ratio (S/N) light curves of the lensed quasars \HEzerozero, \WGzero, \DESzero, \deuxM, \PSseize, and \DESvingt, which were acquired between October 2016 and February 2020 at the MPIA 2.2 m telescope at La Silla. In Sect.~\ref{sec:td}, we detail the time-delay measurement procedure before presenting and discussing our results in Sect.~\ref{section:results}. Our conclusions are summarized in Sect.~\ref{section:conclusion}.
This paper is the second of the TDCOSMO\footnote{\url{www.tdcosmo.org}} series, which includes the COSMOGRAIL\footnote{\url{www.cosmograil.org}}, H0LiCOW\footnote{\url{https://shsuyu.github.io/H0LiCOW/site/}}, STRIDES\footnote{\url{http://strides.astro.ucla.edu}} collaborations, and members of the SHARP collaboration.

%====================
\section{Observation and data reduction}
%====================
\label{sec:data}

\begin{figure}[t!]
    \centering
    \begin{minipage}[c]{0.49\textwidth}
    \includegraphics[width=\textwidth]{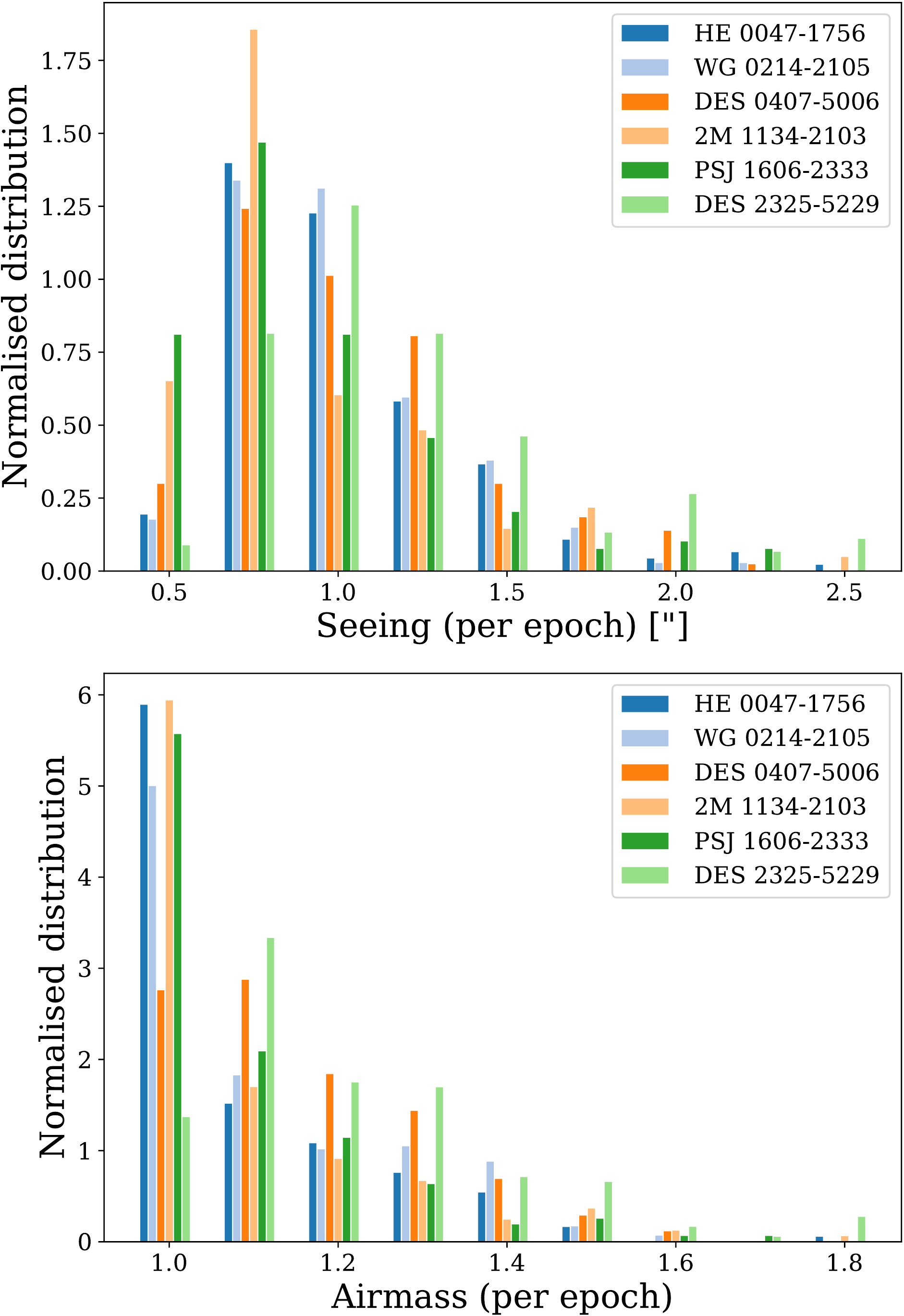}
    \end{minipage} 
    \caption{Seeing and airmass distributions for the six targets monitored with the WFI instrument at the MPIA 2.2 m telescope at ESO La Silla observatory. \label{fig:seeing}} 
\end{figure}

The photometric monitoring data were acquired on a daily basis at the MPIA 2.2 m telescope at ESO La Silla. Each observing epoch consists of four dithered exposures of 320 seconds each, through the $R_c$ filter. The images were taken with the Wide Field Imager (WFI) instrument, which is composed of eight charge-coupled devices (CCD) covering a field of view of $36' \times 36'$ with a pixel size of $0\arcsec238$. A summary of the observing information is presented in Table \ref{tab:lobobs} and Fig.~\ref{fig:seeing}.

 The monitoring campaigns started in October 2016 and ran until February 2020 with a daily planned observing cadence. We observed a total of 11 targets for one full visibility season with the exception of \HEzerozero, which was started in the middle of a season, and \WGzero, for which two seasons were obtained. Among these 11 targets, 9 have sufficiently well-defined features in their light curves to measure the time delays. Three of these targets, namely \DESzeroquatre, \PGonze, and \WFIvingt are presented in previous COSMOGRAIL publications \citep{Courbin2017, Bonvin2018a, Bonvin2019} and 6 are the topic of the present work. The remaining 2, namely \Jzerohuit and \DESvingttrentehuit, will require a second season of monitoring to obtain a robust time delay. These 2 objects are left for future work.
 
 Our data were mainly taken when targets had an airmass below 1.5, but we sometimes relaxed the airmass requirement in order to extend the visibility window. A long seasonal coverage can be crucial in the case of long time delays, when the common features in the light curves only overlap by a few weeks. On average over the six objects presented in this work, one data point per object was recorded every 1.48 d. The actual mean sampling of the light curves is a bit larger than the scheduled daily cadence as a consequence of bad weather and technical maintenance of the telescope. The median seeing over the whole period reported is $1\arcsec06$.

%Added by TeX Support
\begin{table*}[htbp!]
\centering
\caption{Summary of the optical monitoring data in the $R_c$ band. Each epoch consists of 4 exposures of 320 seconds each. The temporal sampling is the mean number of days between two consecutive observations (epochs), excluding the seasonal gap for \HEzerozero and \WGzero. Col. 6 corresponds to the median seeing measured in the images for each object. The seeing and airmass distributions are shown in Fig.~\ref{fig:seeing}. \label{tab:lobobs}}
\resizebox{\textwidth}{!}{
\begin{tabular}{llllcccl}
\hline
Target      & $z_s$ & $z_l$  & Period of observation                     & \#Epochs & Seeing & Sampling & Reference                         \\ \hline \hline
\HEzerozero & 1.66  & 0.407  & Oct. $2^{nd}$ 2016 - Jan. $23^{rd}$ 2018  &    186     &       $1\arcsec09$        &  1.80 days                         & \cite{Wisotzki2004}            \\
\WGzero     & 3.24  & $\sim$0.45 & June $2^{nd}$ 2018 - Feb. $19^{th}$ 2020  & 296     & $1\arcsec08$          & 1.50 days                    & \cite{Agnello2018b} \\
\DESzero    & 1.515 & -      & Aug. $3^{rd}$ 2016 - May $4^{th}$ 2019    & 174     & $1\arcsec09$          & 1.40 days                   & \cite{Anguita2018}             \\
%\Jzerohuit  & 1.116 & 0.659  & Nov. $30^{th}$ 2017 - June $3^{rd}$ 2018  & 143     & $0\arcsec95$          & 1.29 days                  & \cite{Oguri2008}              \\
\deuxM      & 2.77  & -      & Dec. $7^{th}$ 2017 - July $31^{st}$ 2018  & 166     & $0\arcsec92$          & 1.32 days                     & \cite{Lucey2018}               \\
\PSseize    & 1.69  & -      & Jan. $25^{th}$ 2018 - Sep. $23^{rd}$ 2018 & 158     & $0\arcsec95$          & 1.52 days                     & \cite{Lemon2018}              \\ 
\DESvingt    & 2.74  & 0.400      & Apr. $14^{th}$ 2018 - Jan. $6^{th}$ 2019 & 183     & $1\arcsec22$          & 1.33 days                     & \cite{Ostrovski2017}              \\ \hline
TOTAL       & -     & -      & Oct. $2^{nd}$ 2016 - May. $4^{th}$ 2019   & 1163    & -         & -                     & -                                
\end{tabular}}

\end{table*}

The data were reduced according to the standard COSMOGRAIL\footnote{The reduction pipeline can be found at the following address: \url{www.cosmograil.org}} procedure described in detail in \cite{Millon2020}. We first bias-subtracted and flat-fielded the images using sky flats. The sky level was then removed via the {\tt Sextractor} software \citep{Bertin1996}. As the WFI instrument is sometimes affected by fringing in the $R_c$ band, we also constructed a fringe model by iteratively sigma-clipping the four dithered images taken at each epoch and by taking the median. This model was then subtracted from the four individual exposures. 

\begin{figure*}[htbp]
    \centering
    \begin{minipage}[c]{0.49\textwidth}
    \frame{\includegraphics[width=\textwidth]{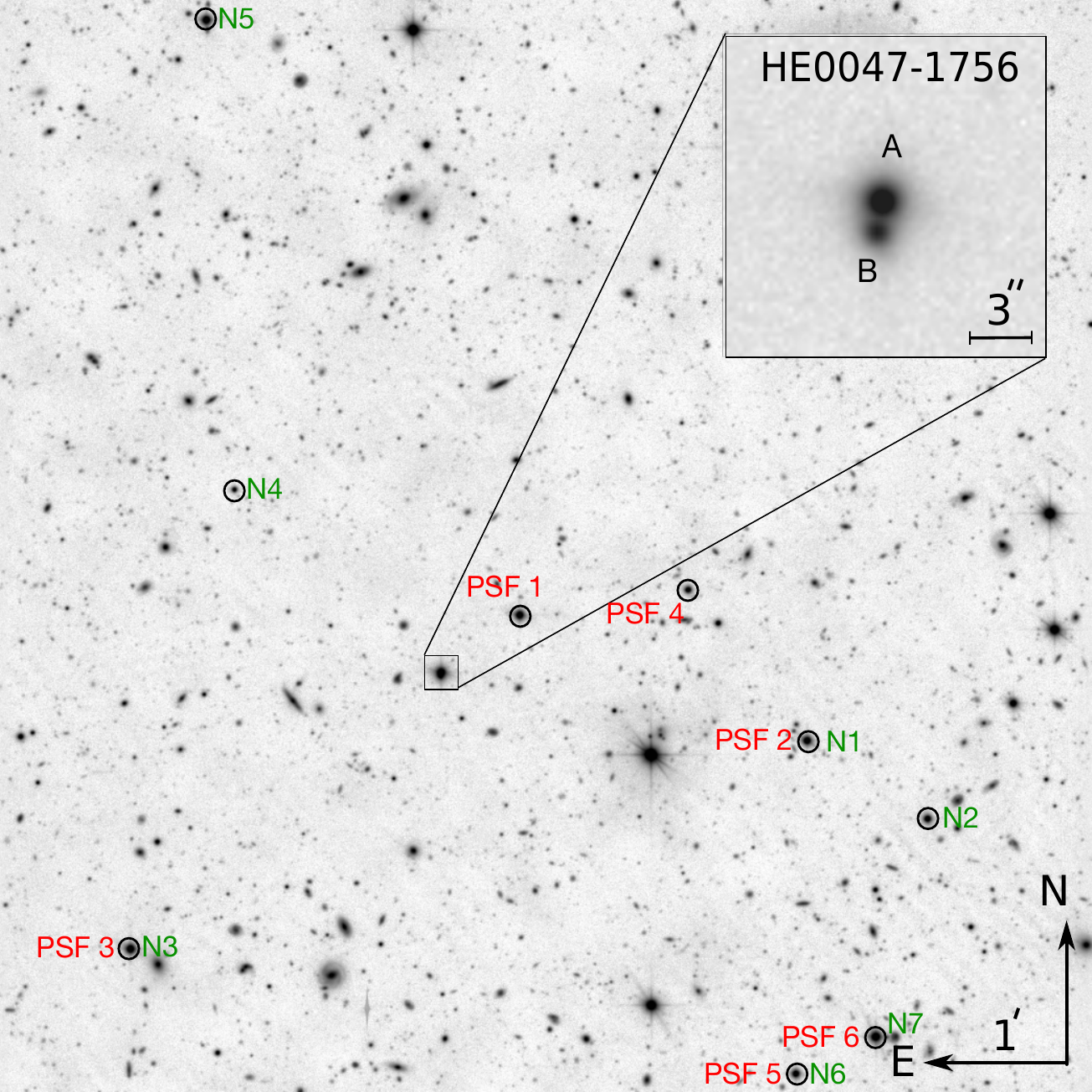}}
    \end{minipage} 
    \begin{minipage}[c]{0.49\textwidth}
    \frame{\includegraphics[width=\textwidth]{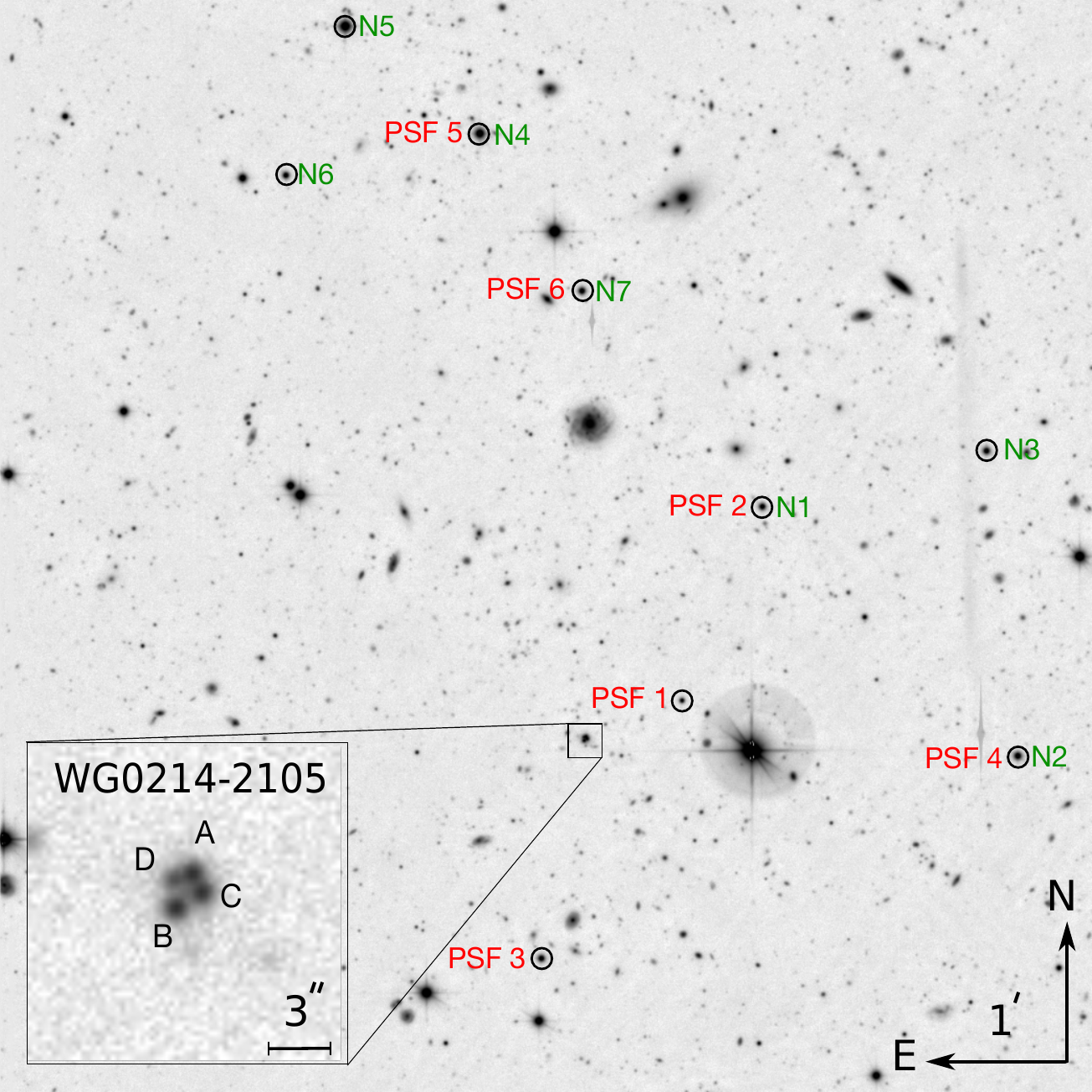}}
    \end{minipage} 
    
    \begin{minipage}[c]{0.49\textwidth}
    \frame{\includegraphics[width=\textwidth]{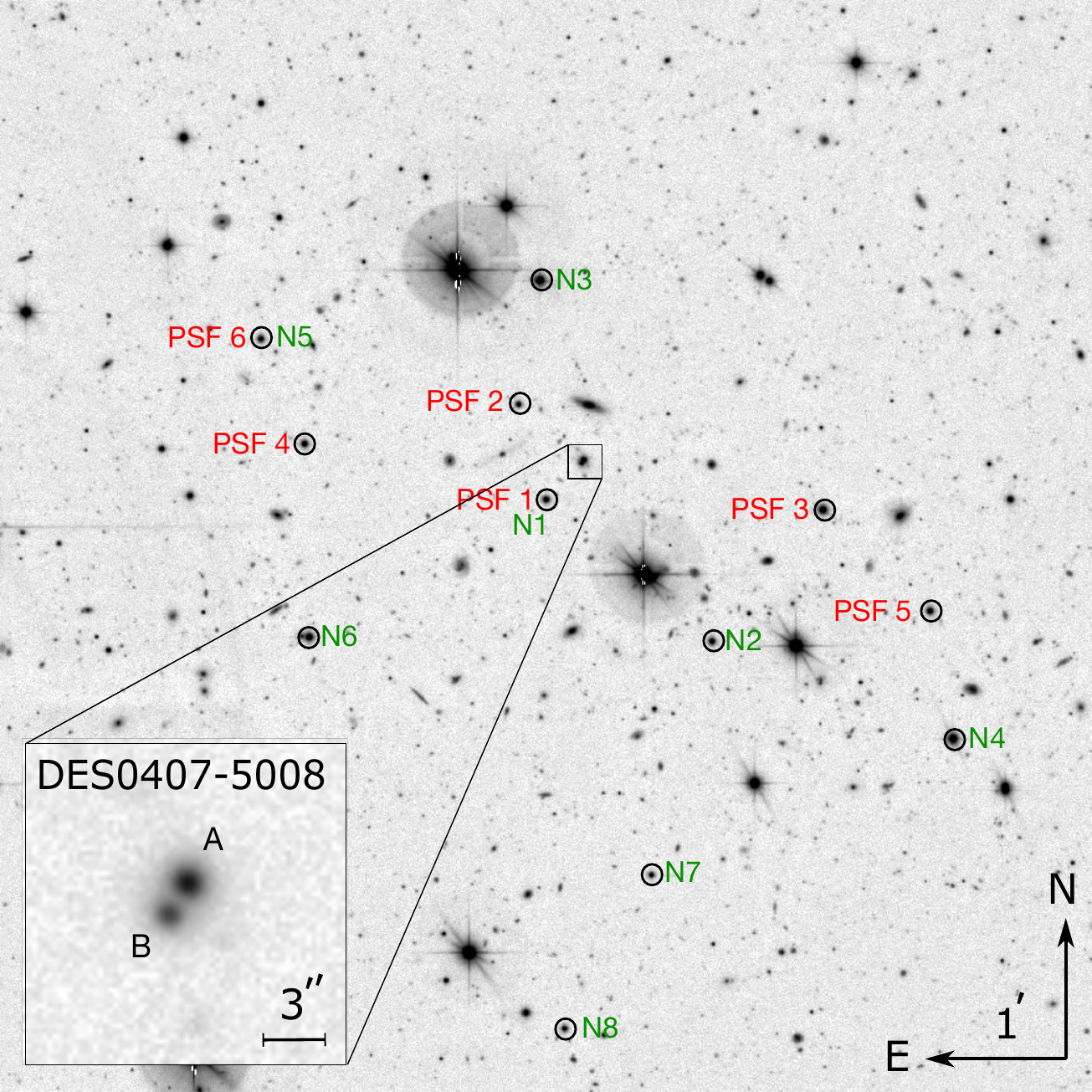}}
    \end{minipage} 
    \begin{minipage}[c]{0.49\textwidth}
    \frame{\includegraphics[width=\textwidth]{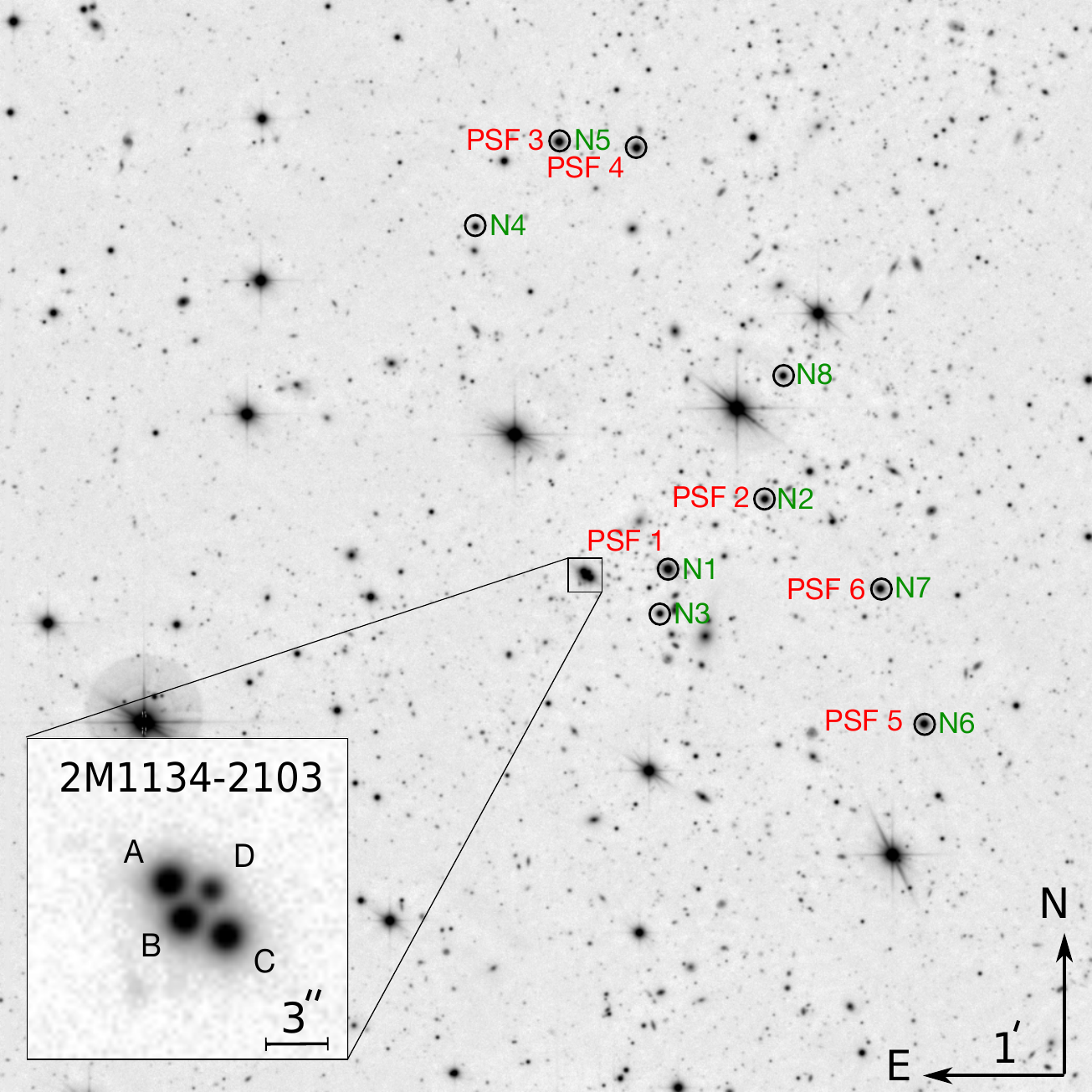}}
    \end{minipage} 

    \caption{Deep stacks of best-seeing images of \HEzerozero (74 images, with a total exposure time of 6.6 h), \WGzero (138 images, 12.3 h), \DESzero(82 images, 7.3 h), and \deuxM (178 images, 15.8 h). The stars used to construct the PSF are circled and labeled in red, whereas the stars used for the night-to-night flux normalization are shown in green. The expanded boxes show single exposures of each lensed quasar in excellent seeing conditions, typically 0\farcs6}
    
    \label{fig:nicefield}
\end{figure*}

\begin{figure*}[htbp]
    \centering
    \begin{minipage}[c]{0.490\textwidth}
    \frame{\includegraphics[width=\textwidth]{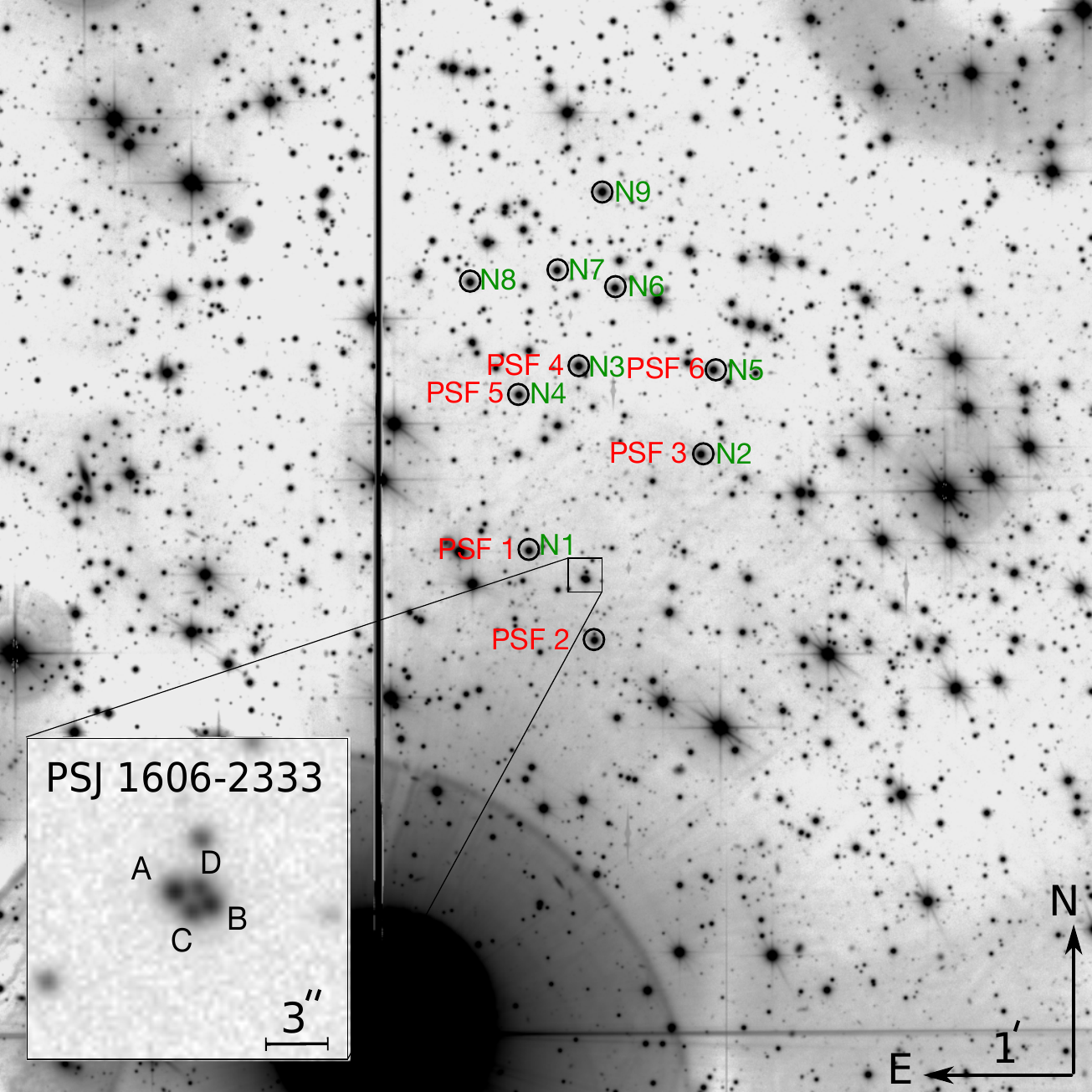}}
    \end{minipage} 
    \begin{minipage}[c]{0.490\textwidth}
    \frame{\includegraphics[width=\textwidth]{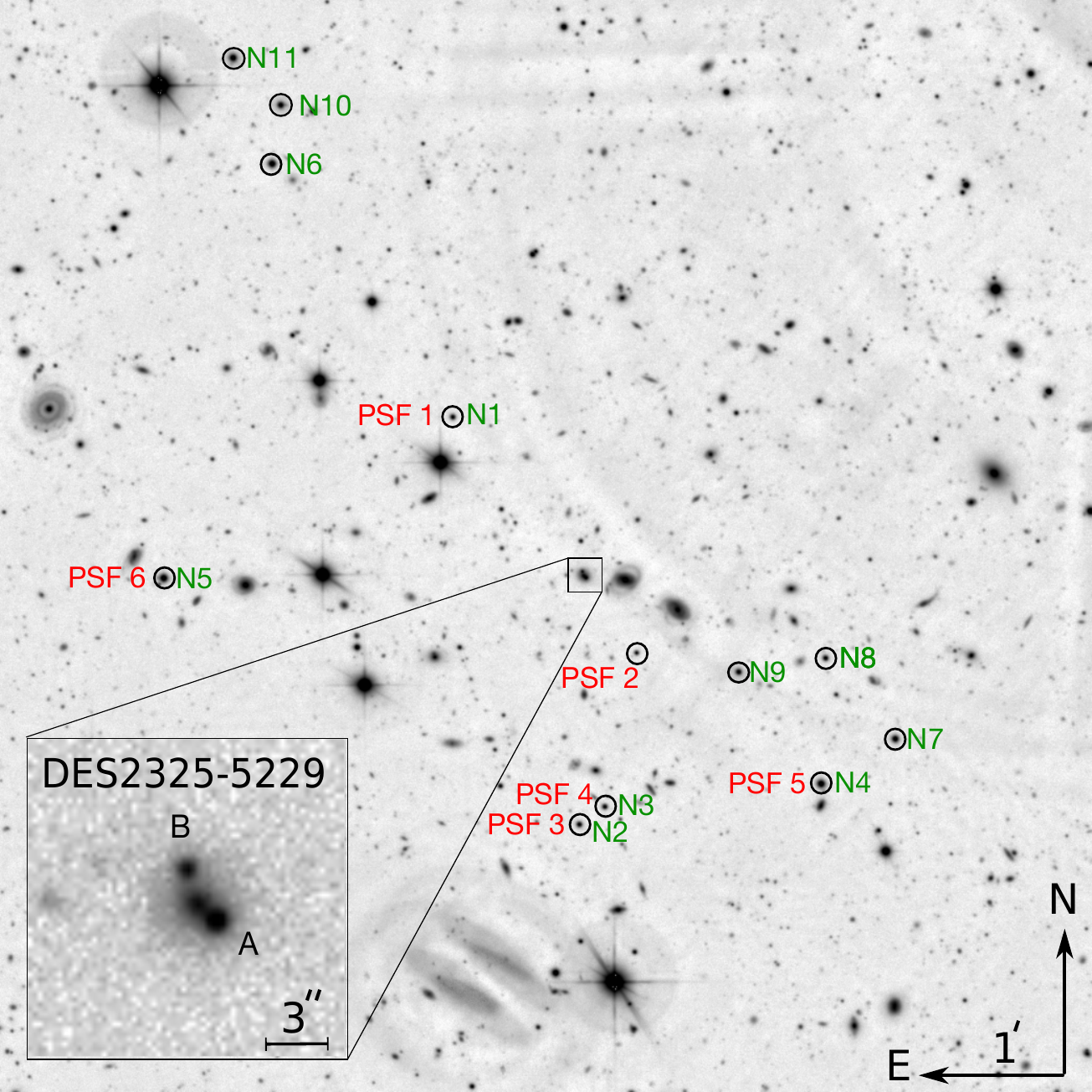}}
    \end{minipage} 
    \caption{Continuity of Fig.~\ref{fig:nicefield} for \PSseize (163 images, 14.5 h) and \DESvingt (184 images, 16.5 h).}
    \label{fig:nicefield2}
\end{figure*}

\begin{figure*}[htbp]
    \centering
    \begin{minipage}[c]{0.49\textwidth}
    \includegraphics[width=\textwidth]{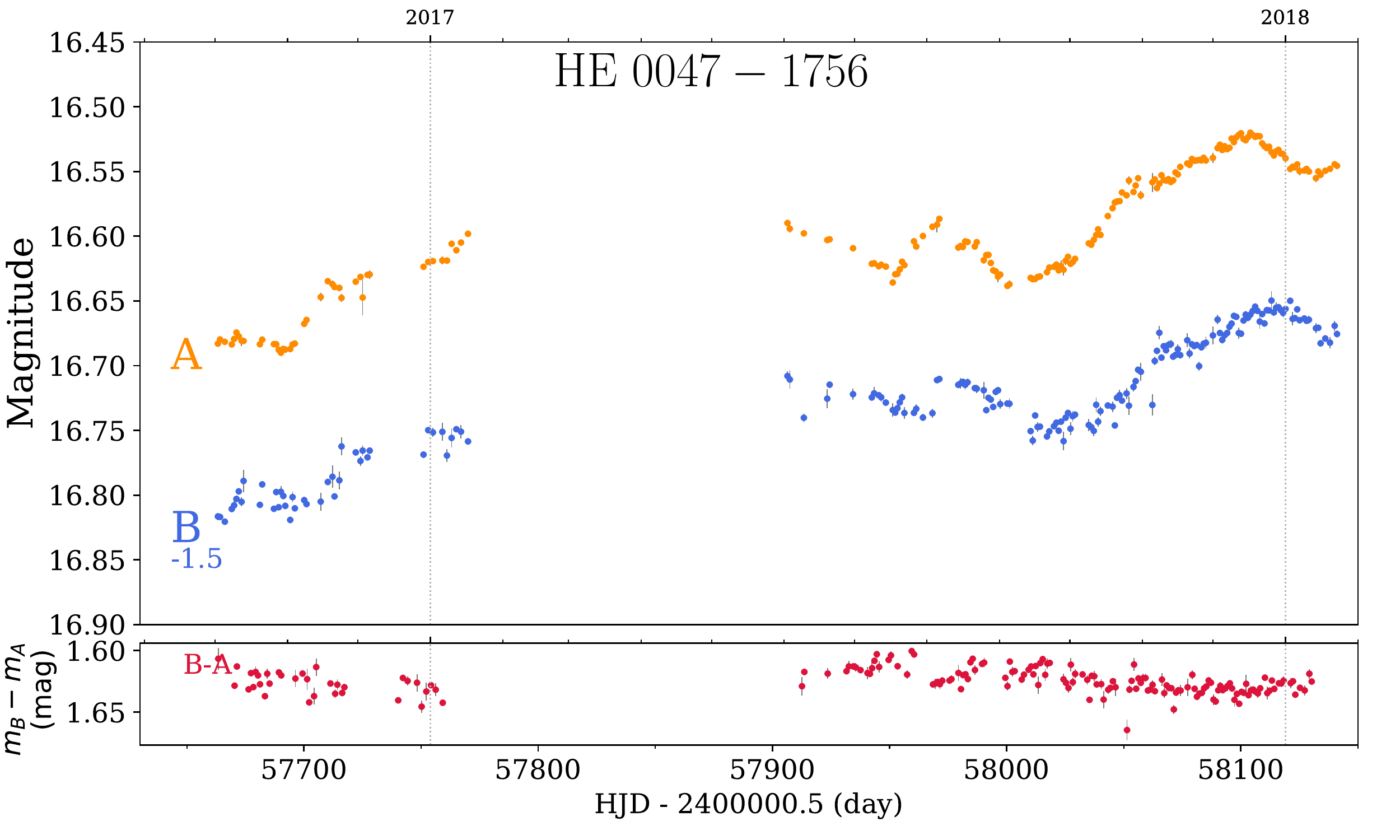}
    \end{minipage} 
    \begin{minipage}[c]{0.49\textwidth}
    \includegraphics[width=\textwidth]{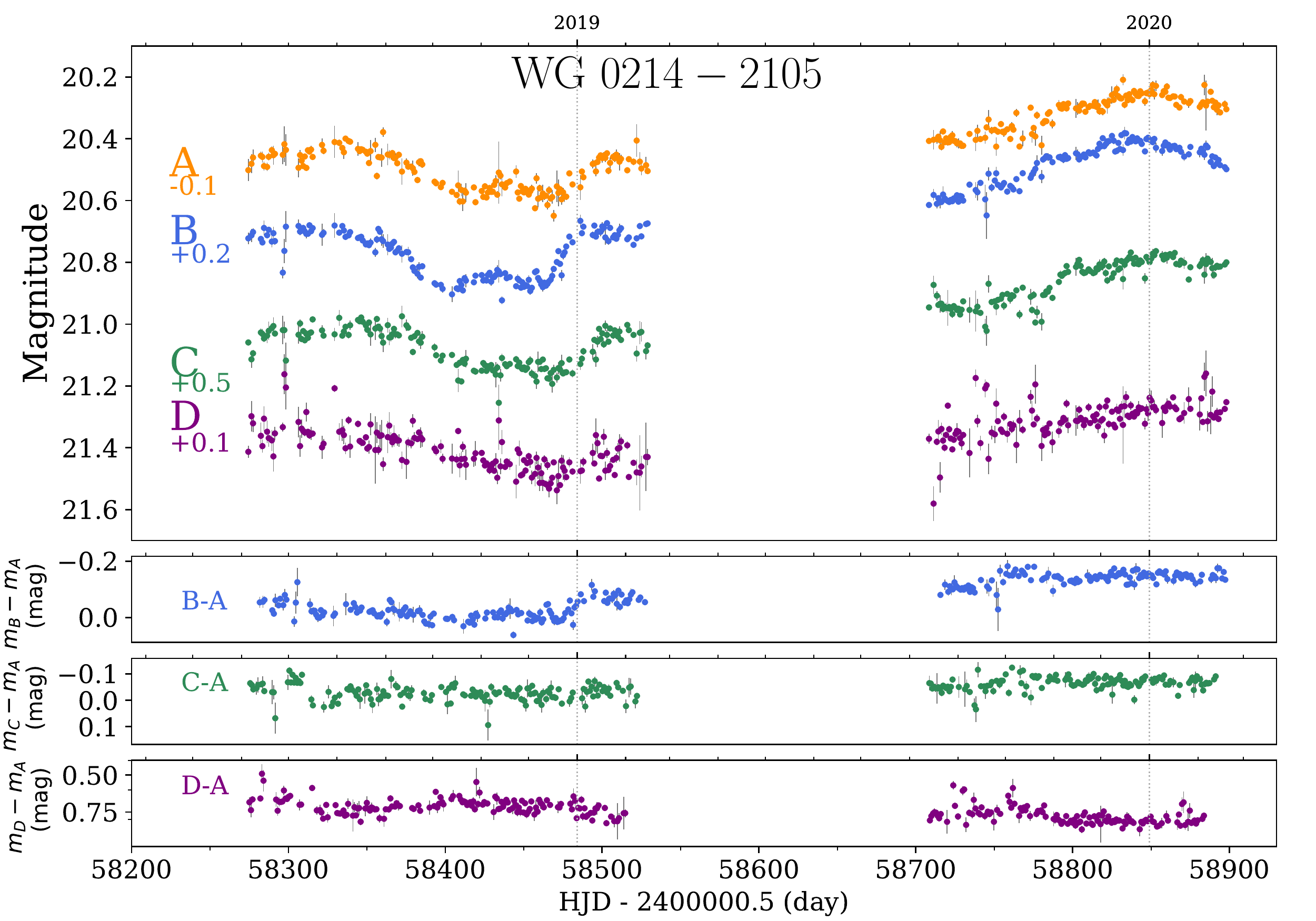}
    \end{minipage} 
    \begin{minipage}[c]{0.49\textwidth}
    \includegraphics[width=\textwidth]{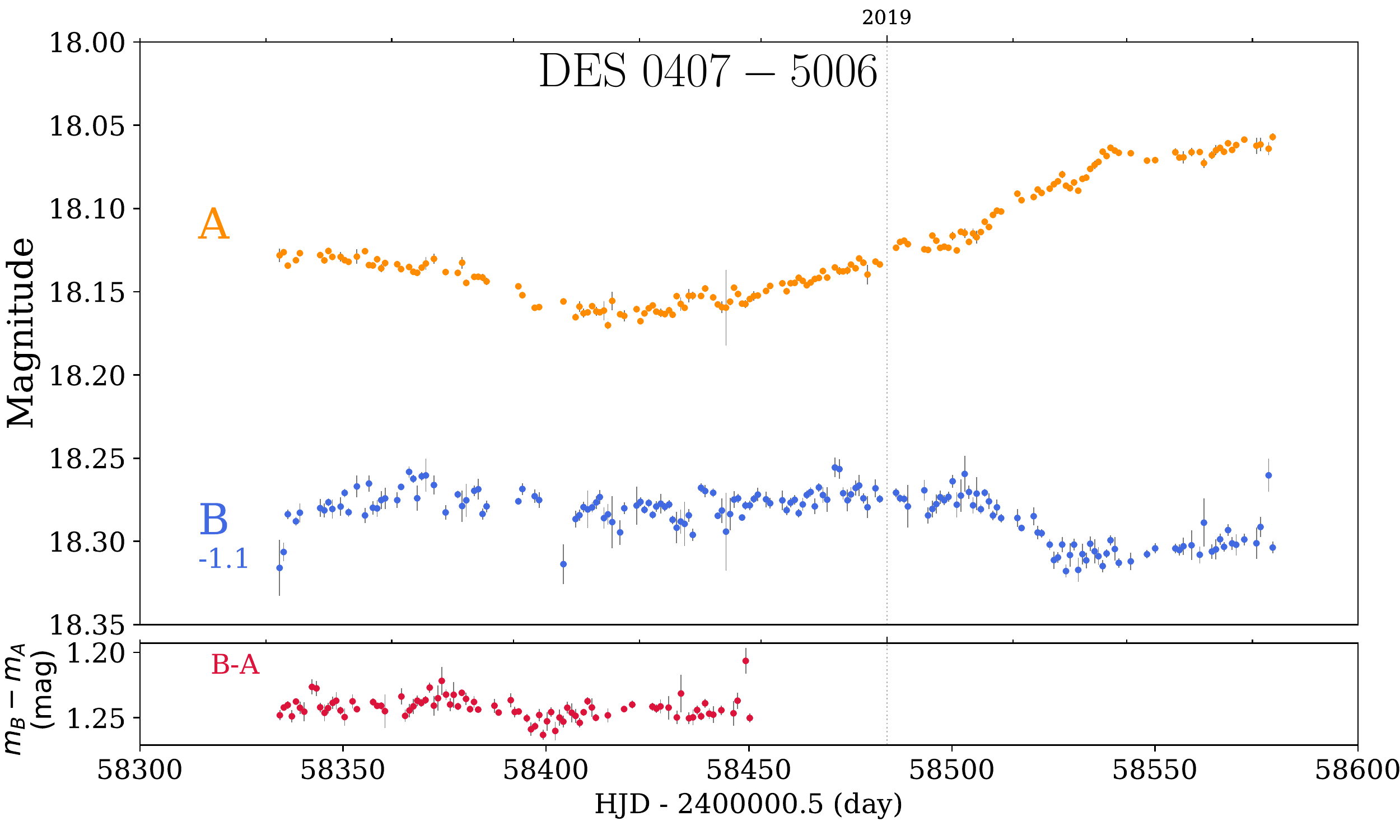}
    \end{minipage} 
    \begin{minipage}[c]{0.49\textwidth}
    \includegraphics[width=\textwidth]{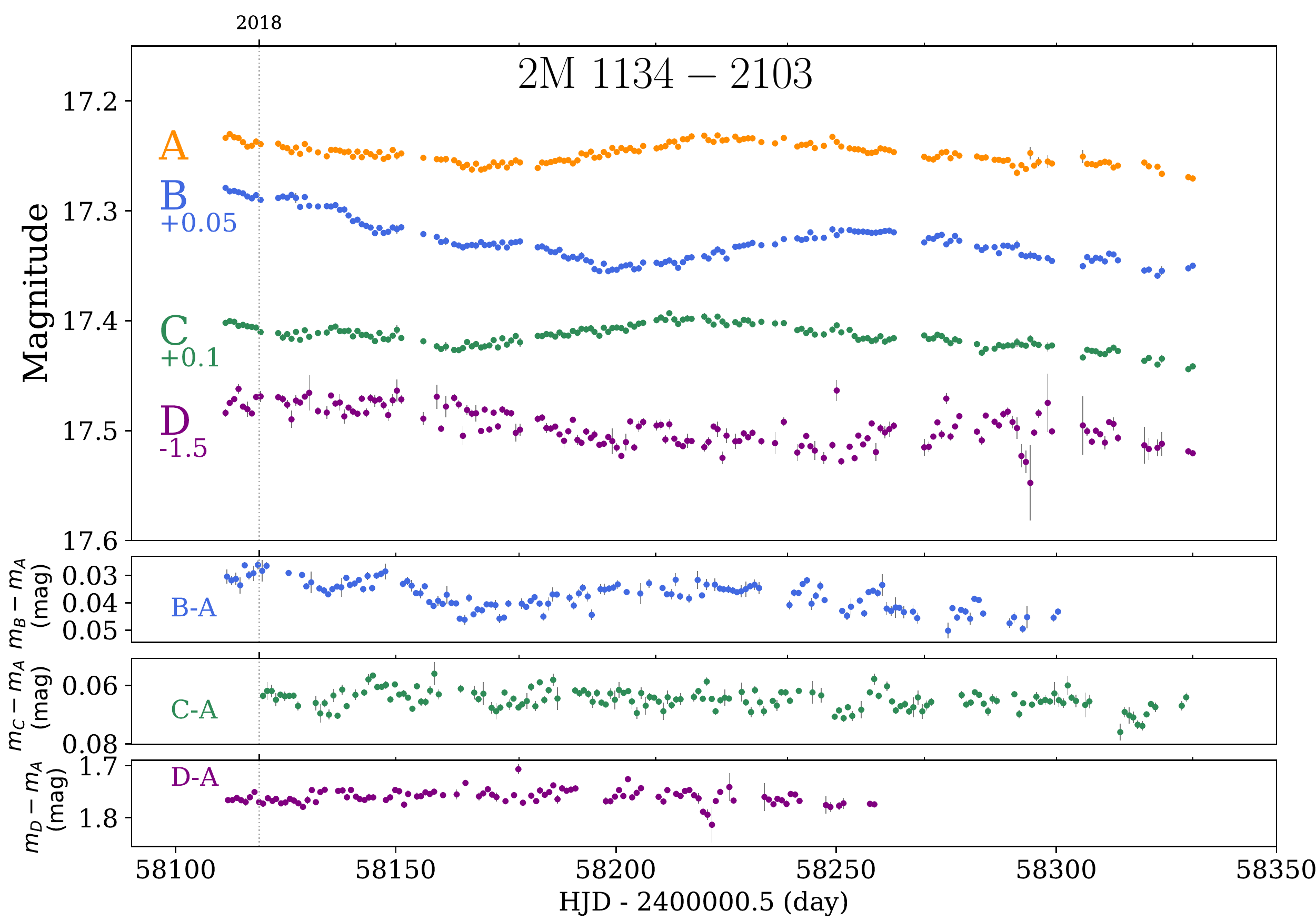}
    \end{minipage} 
    \begin{minipage}[c]{0.49\textwidth}
    \includegraphics[width=\textwidth]{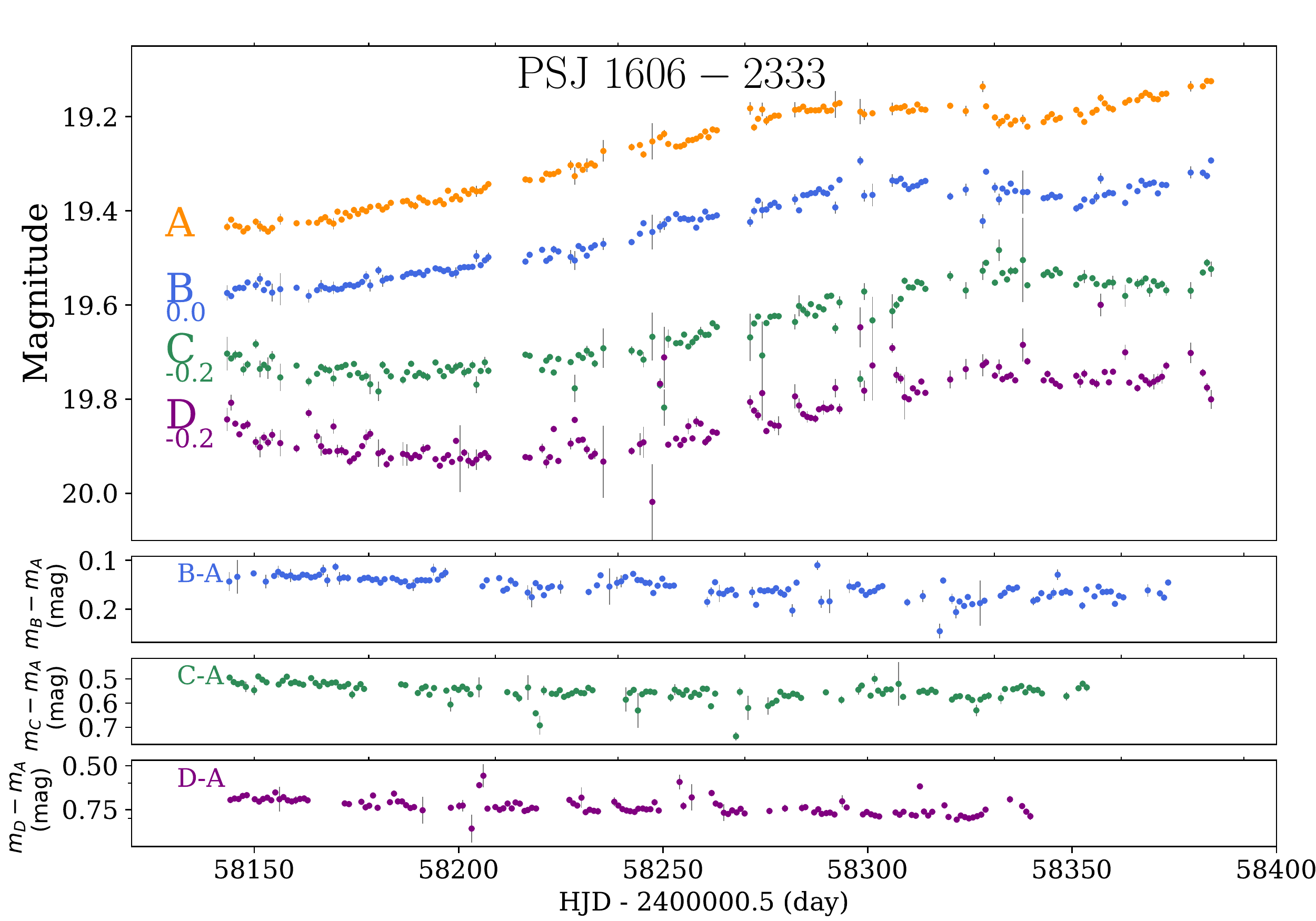}
    \end{minipage} 
    \begin{minipage}[c]{0.49\textwidth}
    \includegraphics[width=\textwidth]{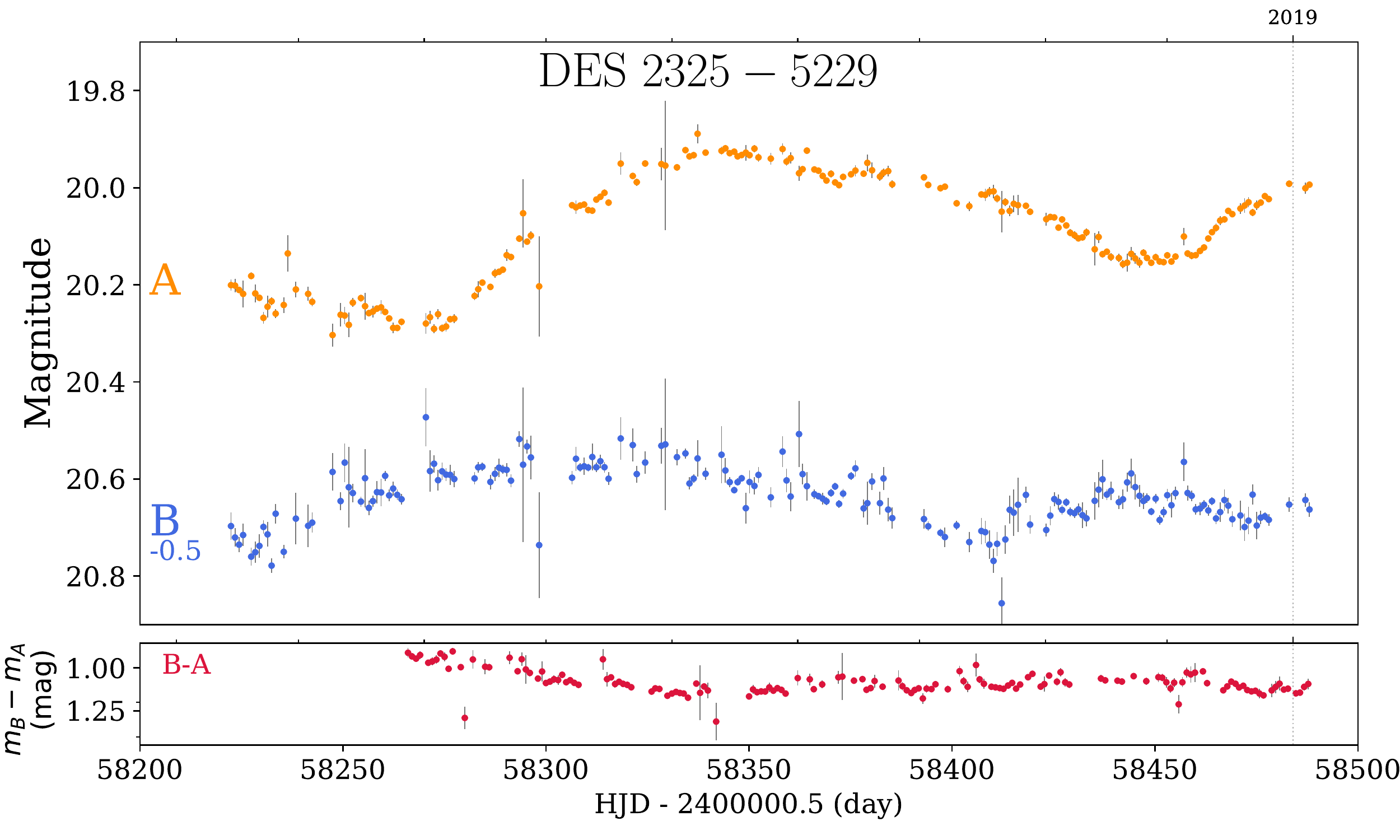}
    \end{minipage} 
    \caption{Light curves for the six lensed quasars presented in this paper. The bottom panels of each lens system show the difference curves between pairs of multiple images shifted by the measured time delays, highlighting the extrinsic variations. Spline interpolation between the data points are used to produce the difference curve, which corresponds the magnitude difference between pairs of images after correction for the measured time delay, but no correction for microlensing is applied.}
    \label{fig:lcs}
\end{figure*}

To obtain an accurate photometric measurement in each single exposure, we performed image deconvolution of the quasar images with the MCS deconvolution algorithm \citep{Magain1998, Cantale2016}. This step largely improves the photometric accuracy as the image separation between multiple images does not exceed a few arcseconds. Fig.~\ref{fig:nicefield} and Fig.~\ref{fig:nicefield2} show the stars used to compute the point spread function (PSF) as well as the reference stars used for image-to-image flux calibration. Each image was deconvolved individually with its own PSF, but all images share the same point source astrometry and the same ``pixel'' channel, which contains all extended sources such as the lensing galaxy, the quasar host galaxy, or companion galaxies \citep[see][for detailed description of the method]{Cantale2016}. The intensities of the point sources are included as free parameters during the process. We computed the median of all individual measurements within a night to produce the light curves presented in Fig.~\ref{fig:lcs}. The photometric error bars for each epoch include the root mean square (rms) standard deviation between the individual measurements as well as systematics due to PSF mismatch during the deconvolution process and normalization errors. These error bars are referred as $\sigma_{\mathrm{emp}}$ in Table \ref{tab:noise}. 

We applied the same deconvolution process to the calibration stars, labeled N1 to NX in Fig.~\ref{fig:nicefield} and Fig.~\ref{fig:nicefield2}, as for the quasar images  to measure their flux. We used the normalization stars for night-to-night calibration relative to a reference image taken in excellent seeing condition. In addition, we used the normalization star labeled N1 for absolute calibration of the light curves. We obtained the corresponding calibrated apparent magnitude in the \emph{r} filter from the PanSTARRS DR2 catalog \citep{PanSTARR}. For the field of \DESvingt and \DESzero, which are not covered by PanSTARRS, we used the \emph{r} magnitude from the Dark Energy Survey (DES) Year-One catalog \citep{DESY1}. These calibrations are only approximate because the \emph{r} filter of DES and PanSTARRS do not exactly match the ESO844 $R_c$ filter used for these observations. 

%====================
\section{Noise properties of the light curves}
%paragraph on data comparison with the DR1 (maybe add the forecast plot by Vivien)
The COSMOGRAIL program was originally designed for monitoring lensed quasars with 1 m-class telescopes and using a biweekly cadence. The photometric precision that can be reached with such instruments in 30 min of exposure per epoch is on the order of 10 mmag rms on the brightest lensed quasars. As a result, only large amplitude variations can be detected. These typically occur on long timescales, on the order of several months or years. Using only the most prominent features of the light curves, it is very difficult to disentangle the intrinsic variations of the quasar from the extrinsic (i.e., microlensing) variations \citep{Bonvin2016, Liao2015} because these extrinsic variations occurs on the same timescale. As a result, it typically requires five to ten seasons of monitoring to obtain enough prominent features in the light curves to unambiguously match the intrinsic variations in the various multiple images without being affected by the extrinsic variations. 

This long-term strategy yielded several precise time-delay measurements \citep{Tewes2013b, Rathna2013, Shalyapin2017, Bonvin2017, Shalyapin2019, Millon2020}, but at a large observational cost. It is no longer sustainable in the era of wide-field surveys such as DES, CFIS, PanSTARRS, and Gaia, which are discovering dozens of new lensed quasars. For example, \citet[][]{Lemon2019, Lemon2018} recently found a total of 46 new lensed quasars by jointly analysing DES, PanSTARRS, and Gaia data. To quickly turn these new systems into cosmological constraints, the time delays must be obtained in just a few seasons. 

The data presented in this work are the result of the high-cadence and high S/N lens monitoring campaigns started in 2016 \citep[see][for the presentation of the program]{Courbin2017}. The enhanced S/N and improved cadence allow us to catch small intrinsic variations of the quasars, which occur on much shorter timescales than typical extrinsic microlensing variations whose timescale ranges from several months to several years \citep[e.g.,][]{Mosquera2011, Millon2020}. In almost all the light curves presented in this paper, intrinsic variations happening on timescales on the order of a few days to weeks can be unambiguously matched in at least the brightest multiple images, making the time-delay measurement possible in one single season.

To emphasize the photometric precision that can be reached in $\sim$20 min exposure per epoch with a 2 m-class telescope, we report in Table~\ref{tab:noise} the noise level in the light curves presented in Fig.~\ref{fig:lcs}. We list the expected median theoretical photon noise from the measured flux $\sigma_{\mathrm{th}}$ and the median empirical noise $\sigma_{\mathrm{emp}}$ obtained from the standard deviation of the measured flux in all four exposures taken in the same night. The quantity $\sigma_{\mathrm{emp}}$ is larger than $\sigma_{\mathrm{th}}$ because it also includes the frame-to-frame normalization errors and the deconvolution errors in addition to the photon noise. We observe that some objects with a wide separation between images and a faint lens galaxy such as \deuxM have almost the same $\sigma_{\mathrm{emp}}$ and $\sigma_{\mathrm{th}}$, which indicates that the photometric errors are still dominated by photon noise and could be reduced by increasing the exposure time. On the contrary, objects with compact image configurations, such as \HEzerozero, seem to be limited by systematic errors possibly introduced by  residual flux contamination after the deconvolution process. Overall, a median empirical photometric precision in the range 1.2-7.1 mmag is reached for at least the brightest quasar image of all lens systems. This allows us to catch intrinsic quasar variation on the order of 10 to 20 mmag in the brightest lens images, which were previously below the noise level of the COSMOGRAIL monitoring campaigns. 

We also present in Table~\ref{tab:noise} the median absolute deviation (MAD) of the residuals after fitting our spline model for extrinsic and intrinsic variations $\sigma_{\mathrm{res}}$ (see Sect. ~\ref{spline} for details). The latter also provides an indication on the smallest intrinsic variations that can be detected by our smooth spline model. This noise estimate is slightly higher than $\sigma_{\mathrm{emp}}$ and $\sigma_{\mathrm{th}}$ because it is impacted by any fast residual variability in the data that cannot be captured by our intrinsic and extrinsic spline models. 

%Added by TeX Support
\begin{table*}
\centering
\small
\caption{\label{tab:noise} Photometric properties of the light curves presented in Fig~\ref{fig:lcs}. We give the maximum image separation in Col. 2; the median observed magnitude over all the epochs in Col. 4; the expected median photon noise, $\sigma_{\mathrm{th}}$, in Col 5; the median empirical photon noise, $\sigma_{\mathrm{emp}}$ in Col. 6 and the MAD of the residuals after fitting our intrinsic and extrinsic spline models in Col. 7. The median expected photon noise, $\sigma_{\mathrm{th}}$, is the theoretical noise expected from the flux counts in the images whereas the empirical photon noise, $\sigma_{\mathrm{emp}}$, corresponds to the standard deviation of the measured flux in the 4 exposures taken in the same night. $\sigma_{\mathrm{emp}}$ corresponds to the photometric uncertainties of the light curves of Fig~\ref{fig:lcs}.}
\begin{tabular}{l|llcccc}
                           & Image separation~             & Image & \multicolumn{1}{c}{\begin{tabular}[c]{@{}c@{}}Magnitude\\~$[$mag$]$\end{tabular}} & \multicolumn{1}{c}{\begin{tabular}[c]{@{}c@{}}$\sigma_{\mathrm{th}}$\\~$[$mmag$]$~\end{tabular}} & \multicolumn{1}{c}{\begin{tabular}[c]{@{}c@{}}$\sigma_{\mathrm{emp}}$~\\~$[$mmag$]$\end{tabular}} & \multicolumn{1}{c}{\begin{tabular}[c]{@{}c@{}}$\sigma_{\mathrm{res}}$ \\~$[$mmag$]$\end{tabular}}  \\ \hline \hline

\HEzerozero & $1\arcsec43$ & A     & 16.6                                & 0.5                                      & 1.5                                                & 2.1                                               \\
                           &                               & B     & 18.22                               & 1.2                                      & 2.4                                                & 4.3                                               \\
\WGzero     & $1\arcsec85$ & A     & 20.53                               & 7.9                                      & 9.8                                                & 12.9                                              \\
                           &                               & B     & 20.48                               & 7.3                                      & 7.1                                                & 10.4                                              \\
                           &                               & C     & 20.5                                & 7.6                                      & 9.1                                                & 11.7                                              \\
                           &                               & D     & 21.26                               & 14.9                                     & 16.3                                               & 21.5                                              \\
\DESzero    & $1\arcsec72$ & A     & 18.13                               & 1.3                                      & 1.7                                                & 2                                                 \\
                           &                               & B     & 19.38                               & 3                                        & 4.1                                                & 4.1                                               \\
\deuxM      & $3\arcsec68$ & A     & 17.25                               & 0.9                                      & 1.2                                                & 1.7                                               \\
                           &                               & B     & 17.28                               & 0.9                                      & 1.4                                                & 1.7                                               \\
                           &                               & C     & 17.31                               & 0.9                                      & 1.4                                                & 1.7                                               \\
                           &                               & D     & 19                                  & 2.4                                      & 3.6                                                & 5.8                                               \\
\PSseize    & $1\arcsec74$ & A     & 19.25                               & 3.1                                      & 4.5                                                & 4.9                                               \\
                           &                               & B     & 19.42                               & 3.4                                      & 4.8                                                & 5.2                                              \\
                           &                               & C     & 19.88                               & 4.8                                      & 6.8                                                & 9.1                                              \\
                           &                               & D     & 20.05                               & 5.5                                      & 8.2                                                & 12.7                                              \\
\DESvingt   & $2\arcsec82$ & A     & 20.07                               & 5.4                                      & 7.2                                                & 9.3                                               \\
                           &                               & B     & 21.14                               & 13                                       & 17.3                                               & 18.3                                            
\end{tabular}
\end{table*}

%====================
\section{Time-delay measurements}
%====================
\label{sec:td}
We used the public Python package \pycs\footnote{PyCS can be downloaded from \url{www.cosmograil.org}}, which contains several algorithms for measuring the time delays in the presence of microlensing \citep{Tewes2013a}. We followed the procedure described in detail in \cite{Millon2020} to robustly measure time delays in an automated way. In doing this, we explored a broad range of choices for our estimator parameters and we estimated the uncertainties on the time delay using simulated light curves containing both the intrinsic and extrinsic variations. We focused on two time-delay estimators, namely the free-knot splines and the regression difference. The free-knot spline estimator was extensively tested on the simulated light curves of the Time Delay Challenge \citep{Bonvin2016, Liao2015, Dobler2015} and showed very good overall performance. 
Throughout this paper, the error bars correspond to 1$\sigma$ uncertainties. Negative A-B time delays means that the variations in image A lead those in image B. 

\subsection{Time-delay measurements with \pycs}
We used the terminology defined by \citet{Bonvin2019}. A curve-shifting technique is a procedure that estimates  time-delay values along with their associated uncertainties given a set of light curves. This technique relies on i) an estimator, which is an algorithm designed to find the optimal time delay between two light curves; ii) estimator parameters, which control the behavior of the estimator; and iii) a generative noise model, which is used to produce simulated light curves, with the same constraining power as the original data. The estimator is also evaluated on simulated light curves to estimate empirically the uncertainties. We briefly describe the two selected estimators in the following section \citep[see][for details]{Millon2020, Tewes2013a}. Estimator parameters used in this work are summarized in Table~\ref{tab:regdiff_param}.

%Added by TeX Support
\begin{table*}
\centering
\caption{Set of parameters used for the regression difference and free-knot spline \pycs estimator. Parameter descriptions can be found in Section \ref{spline} and Section \ref{regdiff}. \label{tab:regdiff_param} }
\begin{tabular}{cc|cccccc}
\multicolumn{2}{c|}{Free-knot splines}                                & \multicolumn{6}{c}{Regression difference}                                                \\ 
\hline\hline
\multirow{3}{*}{ $\eta$ }           & \multirow{3}{*}{15, 25, 35, 45} & \multicolumn{1}{l}{} & Set 1 & Set 2   & Set 3   & Set 4   & Set 5      \\ 
\cline{4-8}
                                    &                                 & $\nu$                & 1.7                    & 1.8     & 1.3     & 1.5     & 1.9        \\
                                    &                                 & A                    & 0.5                    & 0.6     & 0.3     & 0.4     & 0.7        \\
\multirow{3}{*}{$n_{\mathrm{ml}}$ } & \multirow{3}{*}{0,1}           & scale                & 200                    & 150     & 150     & 250     & 250        \\
                                    &                                 & errscale             & 20                     & 15      & 10      & 25      & 25         \\
                                    &                                 & kernel               & Mat\'ern                & Mat\'ern & Mat\'ern & Mat\'ern & Power Exponential 
\end{tabular}
\end{table*}

\subsubsection{Free-knot spline estimator}
\label{spline}
This estimator relies on the construction of an ``intrinsic'' model to represent the quasar variations common to all the light curves up to a time and magnitude shift and an ``extrinsic'' model to represent the additional sources of variability that differ between the light curves. This typically includes variability introduced by the stars in the lens galaxy. Both models use free-knot B-spline to fit the light curves \citep{molinari2004bounded}. The algorithm simultaneously optimizes the position of the knots of the intrinsic and extrinsic splines as well as the time delays and magnitude shifts between the light curves. 
The flexibility of the fit is controlled by two estimator parameters. The first, $\eta$, corresponds to the initial mean spacing between knots of the intrinsic spline and the second, $n_{ml}$, corresponds to the number of internal nodes for the extrinsic splines per observing season, equally distributed over the monitoring period. When we have only one season of monitoring per object, we fix the knot position of the extrinsic splines to avoid introducing too much freedom into the microlensing models as the latter are not expected to vary on timescales shorter than a few weeks. We note that $n_{ml}=0$ means that the extrinsic splines contain only two knots at each extremity of the light curves and therefore correspond to polynomials of degree 3.

\subsubsection{Regression differences estimator}
\label{regdiff}
This second method first performs a regression with Gaussian processes on each light curve individually. The regressions are then shifted in time and subtracted pair-wise. The algorithm optimizes the time shift between the curves by minimizing the variability in the subtracted light curve. This approach does not explicitly model  the extrinsic variations and is therefore fundamentally different from the free-knot splines method. This estimator also relies on a choice of parameters to control the smoothness of the fit with Gaussian processes. Consequently, the kernel function of the Gaussian process,  its smoothness degree, $\nu$, its amplitude, A, its scale, and an additional scaling factor of the photometric errors need to be adjusted. We tested five different sets of parameters that visually provide a good fit of the data. 

For each estimator and estimator parameters, we ran the optimization 500 times from different starting points (i.e., guess time delay) on the same observed light curves. This is meant to ensure that the time-delay estimator has converged and that a robust time-delay estimate can be measured independently of the initial guess for the time delay. We took the median value of the distribution as our central time-delay estimate. This procedure is not a Monte Carlo approach and we do not use the standard deviation of the distribution as our final uncertainties. The procedure to measure the uncertainties requires the generation of simulated light curves and is summarized below. 

\subsection{Uncertainties estimation on the time delay with \pycs}
In \pycs, the uncertainties are estimated in an empirical way, by generating simulated light curves that have similar constraining power as the original data. These simulated curves are identical to the data in terms of temporal sampling, intrinsic variations of the quasar, and extrinsic variations. We used the same intrinsic and extrinsic splines to generate all simulated light curves. However, they differ from the real data in their time delays and their realization of correlated and Gaussian photometric noise. For each set of estimator parameters, a generative noise model produces 800 different realizations of the curves that statistically match the observed data in terms of correlated and Gaussian noise. The true time delays encoded in the simulated curves are in the range $\pm$10 days around our initial estimation obtained by running the estimator on the real data. We followed this procedure using the automated version of \pycs described in detail in \cite{Millon2020}. 

The estimators were run on the simulated light curves and we obtained the final uncertainties for a given curve-shifting technique (i.e., an estimator, a set of estimator parameter, and a generative noise model) by adding in quadrature the systematic and random errors between the measured and true time delays.

\subsection{Combining the curve shifting techniques}
\label{section:comb_estimate}
To combine the curve shifting techniques and obtain our final time-delay estimates for each object, we first combined the curve-shifting techniques that share the same estimator, that is, the regression difference or the free-knot spline, which have different sets of estimator parameters.
The marginalization over the model parameters cannot be done in a fully Bayesian framework, as this would require a very large amount of computation to properly sample the parameter space. To keep the computation time manageable on a small-scale computing cluster, we prefer to probe the parameter space in a grid-wise fashion. The explored parameter space is limited to a region that provides reasonable uncertainties, indicating a good fit quality.

In addition, we cannot use the $\chi^2$ or any derived model selection criteria (e.g., the Bayesian information criterion (BIC) or the Akaike information criterion (AIC)) to estimate the weight of each model due to the degeneracy between intrinsic and extrinsic variations. Because of this degeneracy, it is not possible to define a proper metric to quantify the quality of the fit. We therefore prefer to apply the same methodology as first introduced in \cite{Bonvin2018a}. The goal of this method is to obtain a trade-off between an optimization and a marginalization over the estimator parameters. A pure optimization selects the set of estimator parameters that gives the most precise time-delay measurement, but the price to pay is neglecting all the other models for the quasar variability and extrinsic variations that are not necessarily compatible within statistical uncertainties. On the other hand, marginalizing over all estimator parameters unnecessarily increases the uncertainties as all models are not equally plausible and do not yield the same fit quality. 

To solve this problem, \cite{Bonvin2018a} proposed to first select the most precise estimate as a reference and to compute its tension, $\tau$, with all other estimates. If the tension exceeds a certain threshold $\tau_{\rm thresh} = 0.5$, we combine the most discrepant estimate with the reference. This combined estimate becomes the new reference and we repeated this process until no further tension exceeds $\tau_{\rm thresh}$. We also checked that the choice of $\tau_{\rm thresh}$ did not significantly change the final estimate. We note that choosing $\tau_{\rm thresh} = 0$ corresponds to a marginalization between all the available sets of estimator parameters, whereas choosing $\tau_{\rm thresh} = +\infty $ selects only the most precise set.

\begin{figure}[ht!]
    \centering
    \includegraphics[width=0.49\textwidth]{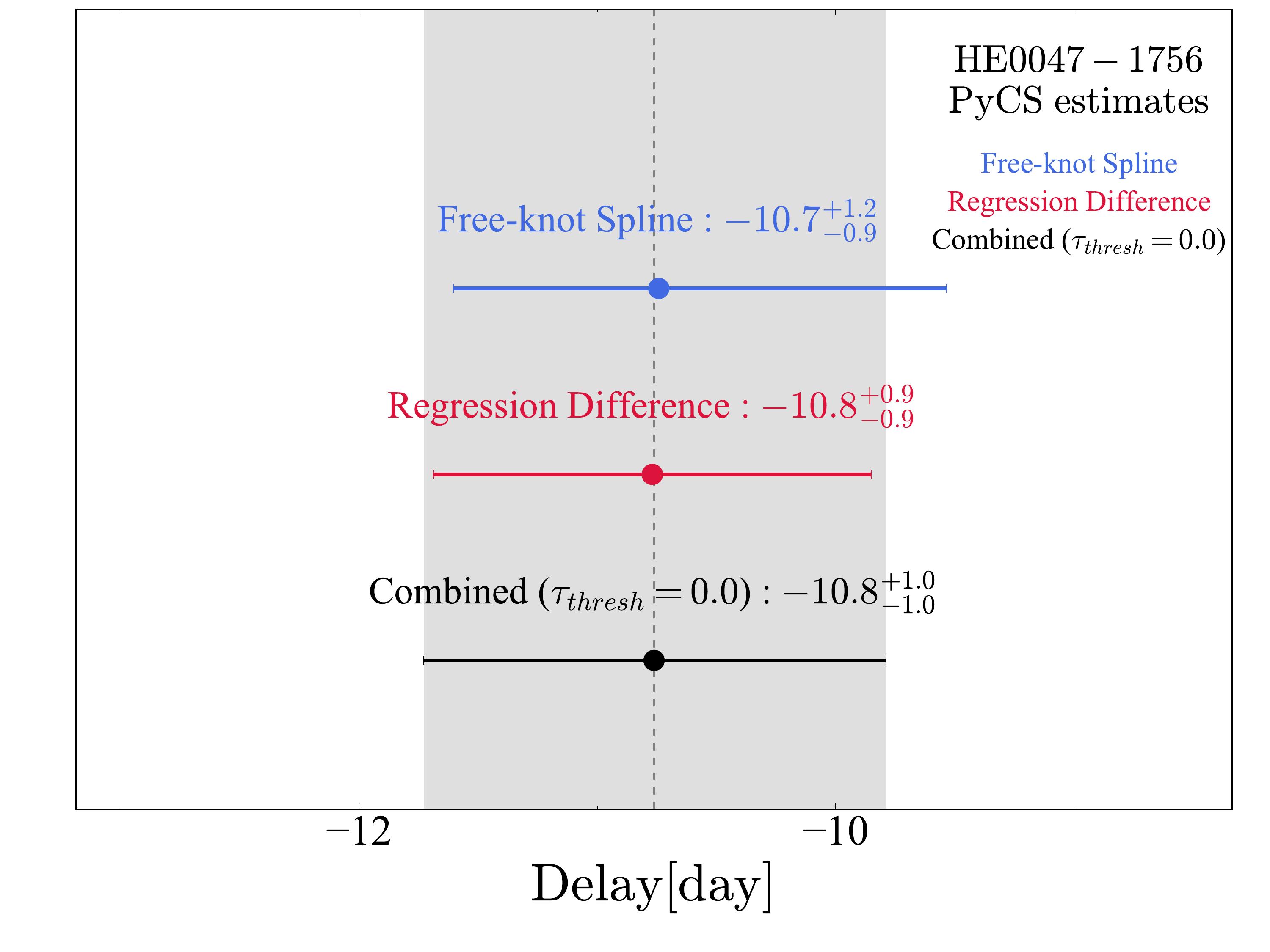}
    \includegraphics[width=0.49\textwidth]{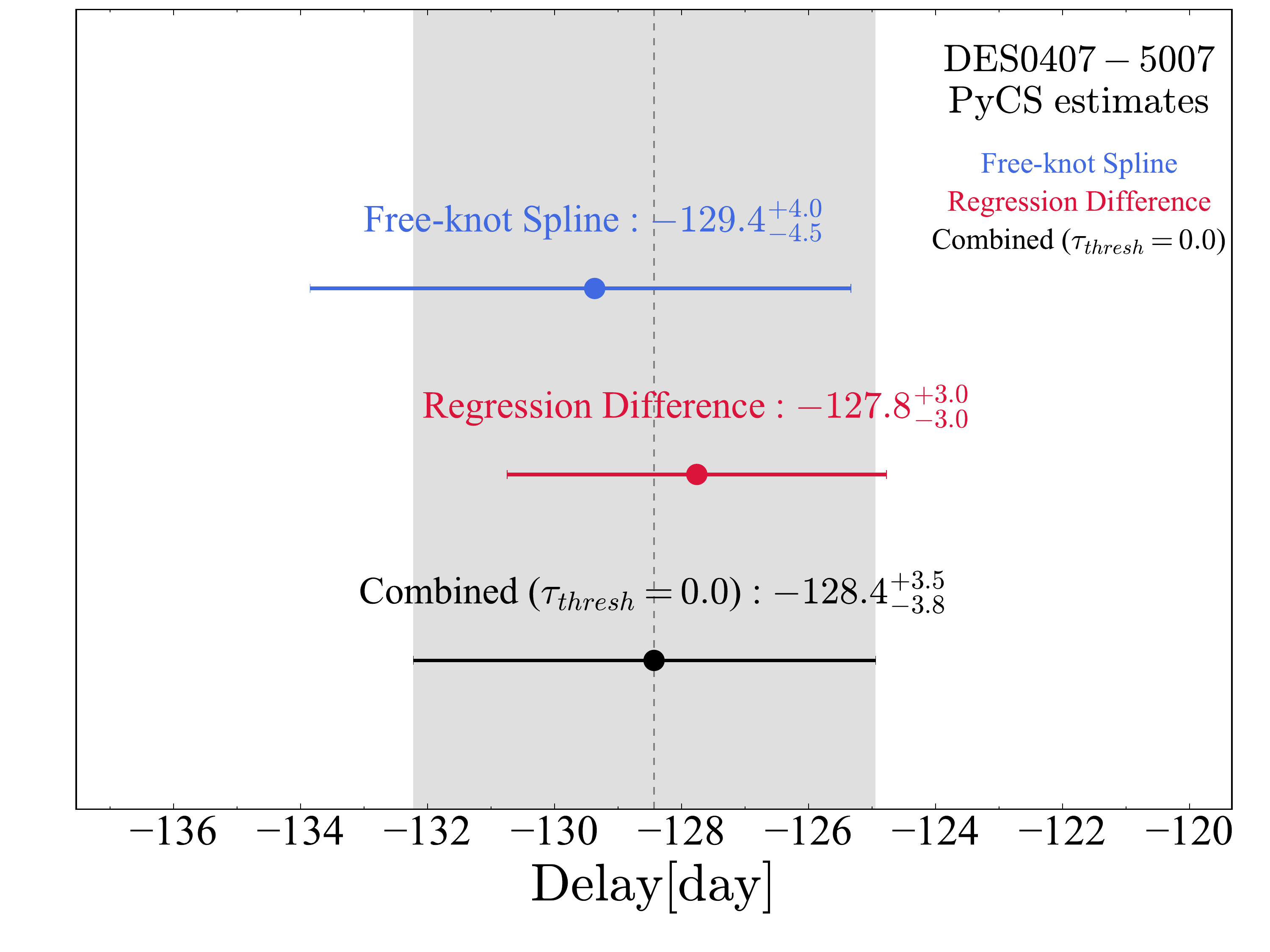}
    \includegraphics[width=0.49\textwidth]{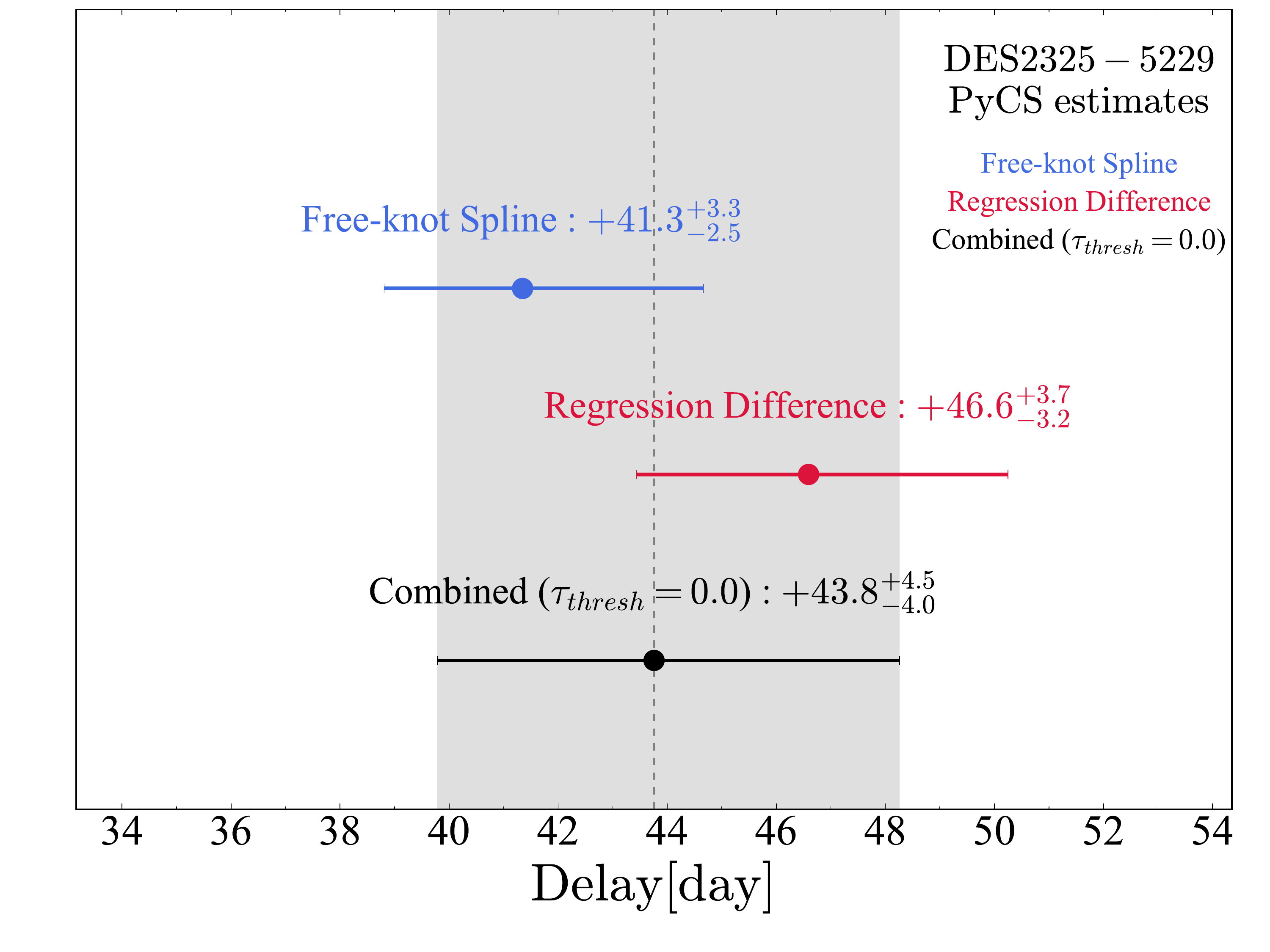}
    \caption{Time-delay estimates for \HEzerozero, \DESzero, and \DESvingt measured with the regression difference estimator (in red) and with the free-knot spline estimator (in blue). The marginalization over the two estimators is shown in black.}
    \label{fig:delay0}
\end{figure}

\begin{figure*}[htbp!]
    \centering
    \includegraphics[width=0.95\textwidth]{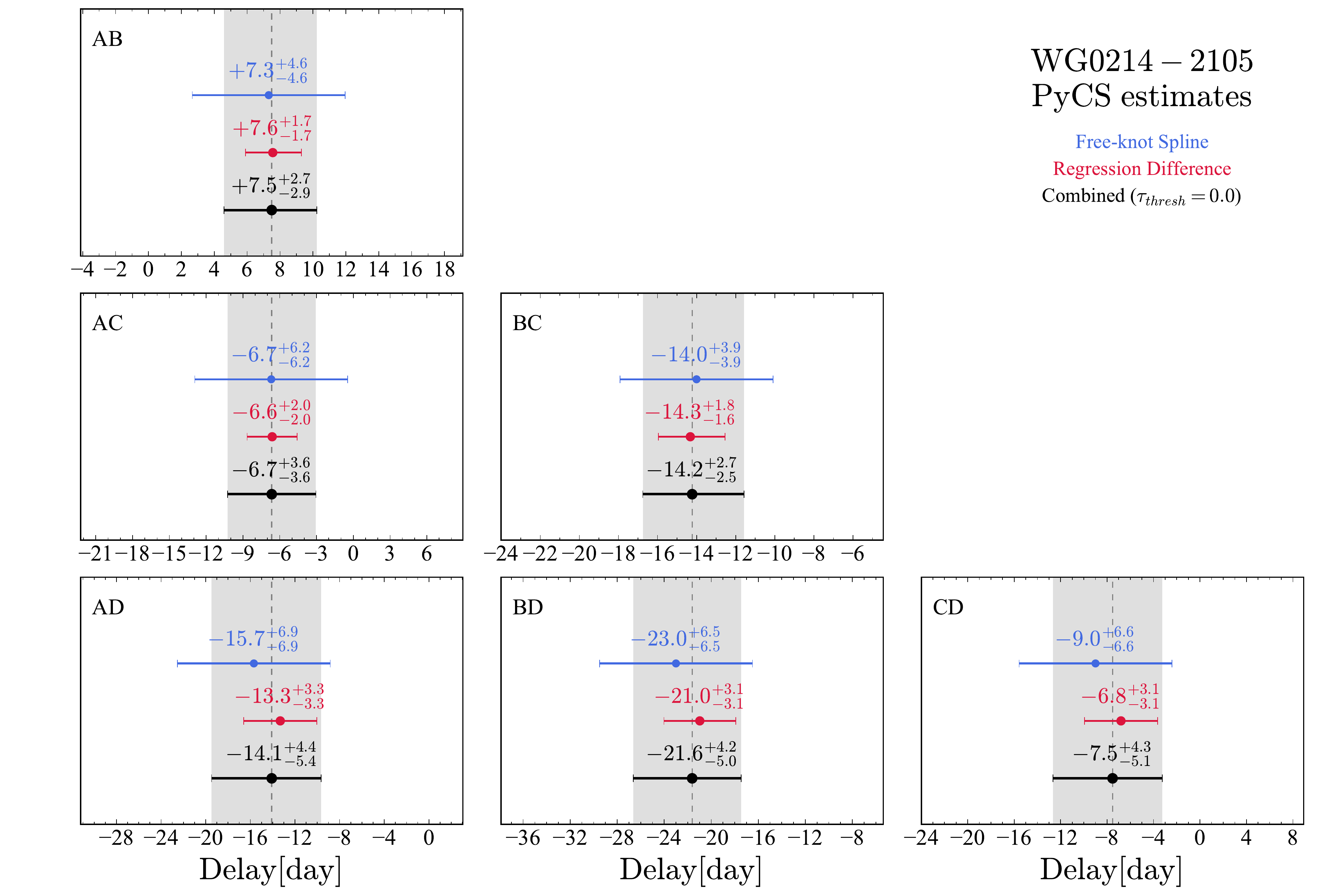}
    \caption{Same as Fig.~\ref{fig:delay0} for \WGzero.}
    \label{fig:delay1}
\end{figure*}

\begin{figure*}[htbp!]
    \centering
    \includegraphics[width=0.95\textwidth]{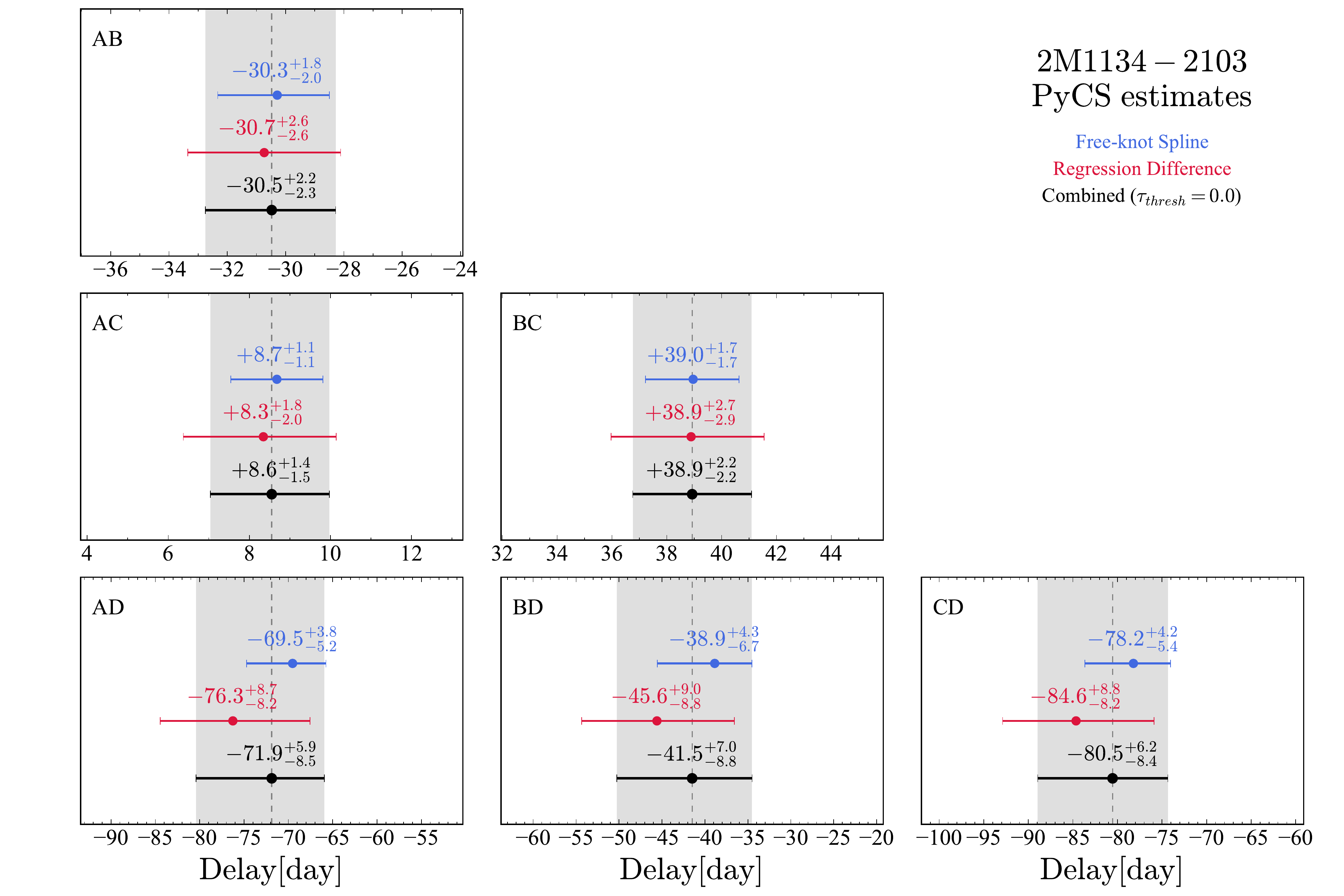}
    \includegraphics[width=0.95\textwidth]{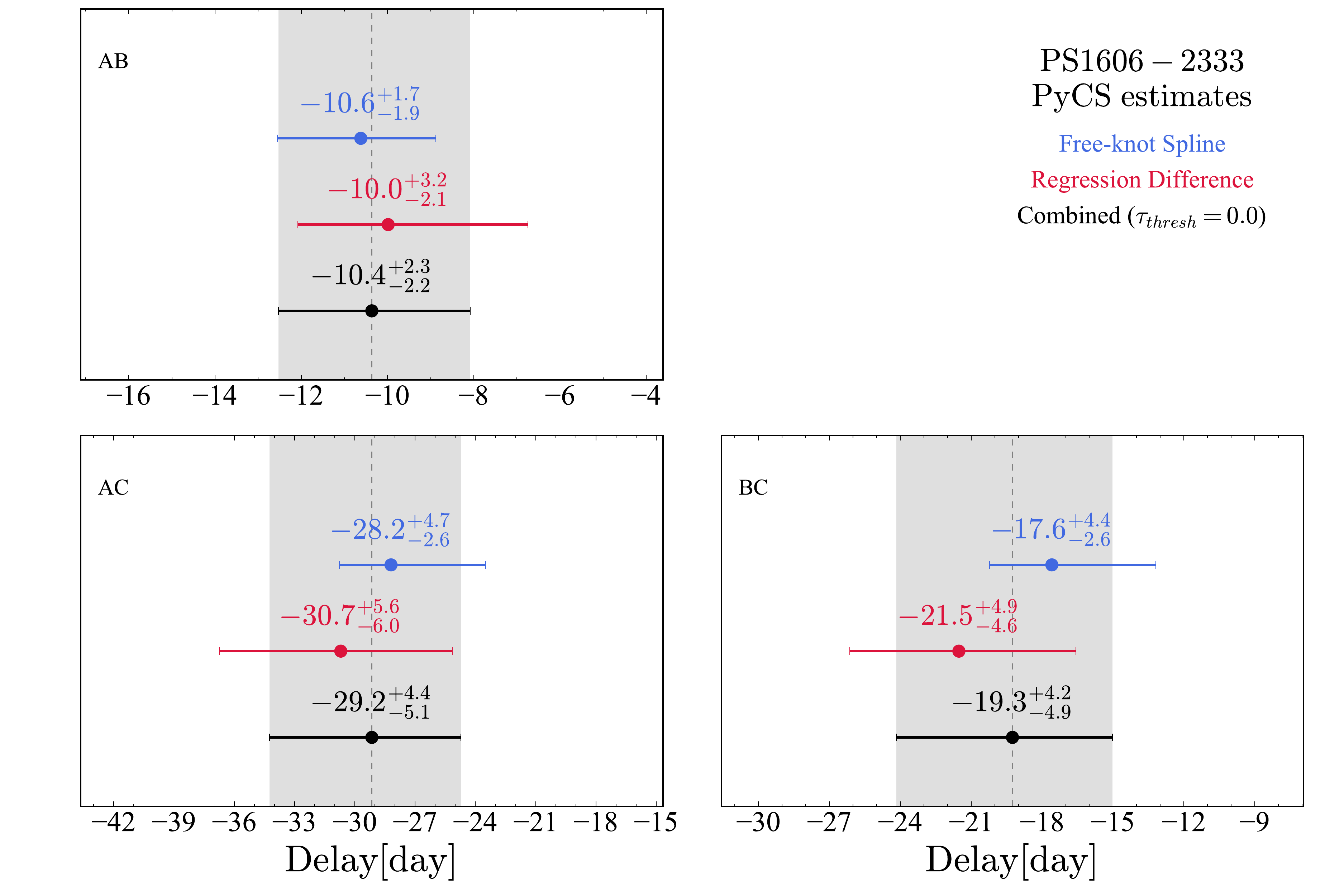} %contain ml only 
    \caption{Same as Fig.~\ref{fig:delay0} and Fig.~\ref{fig:delay1} for \deuxM and \PSseize.}
    \label{fig:delay2}
\end{figure*}

We obtained our final time-delay estimates for each pair of light curves and for each estimator by applying this procedure on the data. These results are presented in Fig.~\ref{fig:delay0}, Fig.~\ref{fig:delay1}, and Fig.~\ref{fig:delay2}. As the two estimators are intrinsically different but are applied to the same data set, they can not be considered as two independent measurements of the time delays. We therefore propose a marginalized estimate over the two curve shifting algorithms. These are shown in black in these same figures. 

%\subsection{Combining with other data sets}
%
%For one object, namely \HEzerozero, we improve the precision of our measurements by combining with the time delays measured using the decade-long light curves previously obtained by the COSMOGRAIL program \citep{Millon2020}. As these two sets are fundamentally different in term of data quality and do not cover the same period, we consider these two experiments to be independent and therefore combine the two time delay estimates in a single “PyCS-mult" estimate. 

%\HEzerozero was previously monitored at the Swiss Leonhard Euler 1.2m (Euler) telescope from 2004 to 2010 with the C2 instrument and from 2011 to 2018 with the ECAM instrument. The final “PyCS-mult” estimate is therefore based on the combination of these two data sets together with the present new WFI data set. The results are presented in Fig.~\ref{fig:delay_combi}.

%====================
\section{Results}
\begin{figure*}
\centering
\hspace{1cm}
\includegraphics[width=0.8 \textwidth]{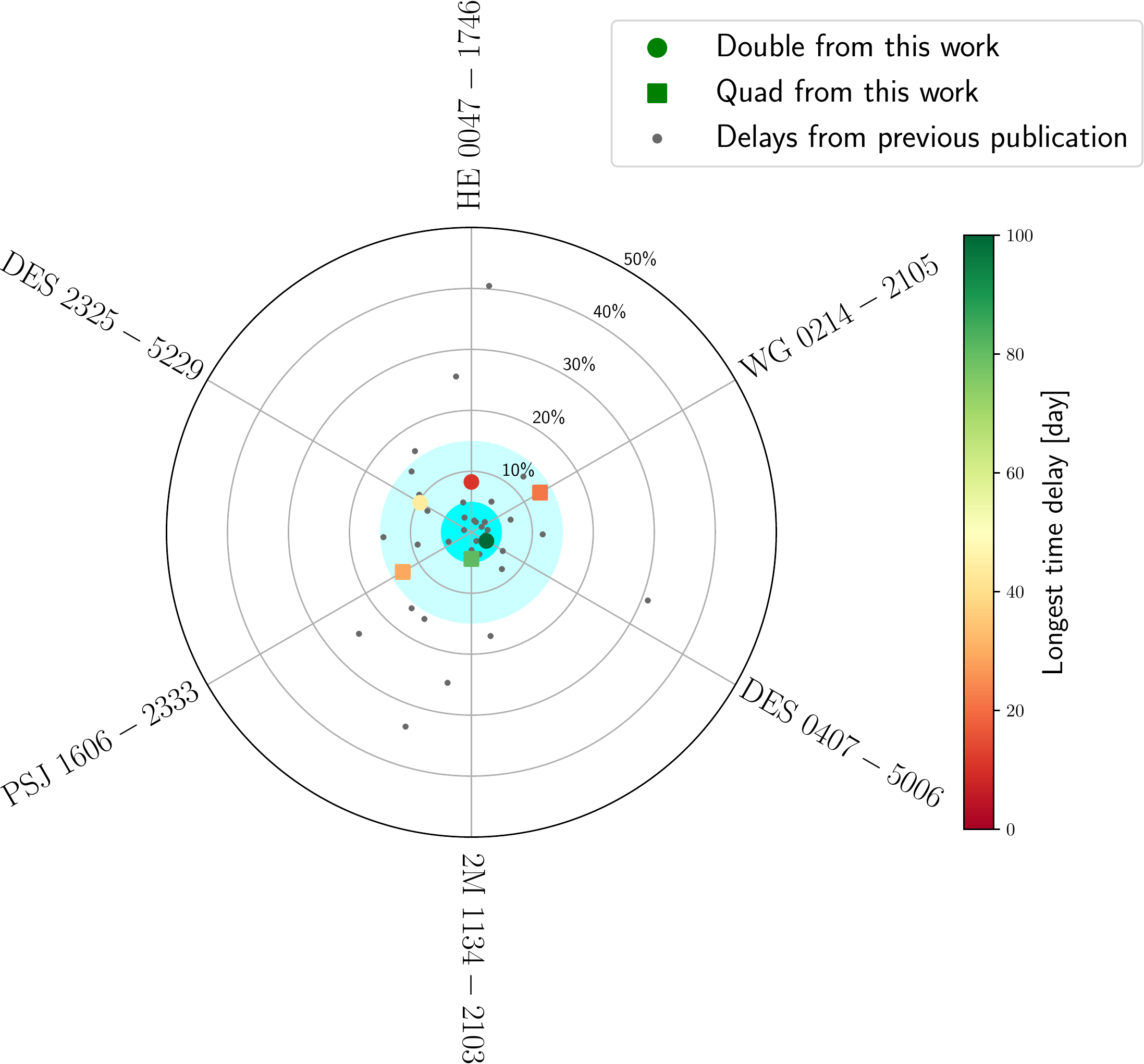}
\caption{\label{fig:delayplot} Time-delays relative uncertainties for the object presented in this work (colored dots) and already available in the literature (gray dots) \citep[see Table 3 of][for a list of published time delays]{Millon2020}. For quadruply imaged quasars,  the combined uncertainties between all three independent time delays, corresponding to the minimal uncertainties achievable on \hc, are shown under the assumption that the time-delay errors remain the dominant source of uncertainties. The outer light blue circle corresponds to a precision better than 15\%. The inner blue circle corresponds to the target region with precision better than 5\%, corresponding to the threshold at which the time-delay errors become smaller than other sources of errors in the inference of \hc.}
\end{figure*}

%====================
\label{section:results}
The procedure described in Sect.~\ref{sec:td} was applied to the six lensed quasars presented in this paper.  Table~\ref{tab:tdelays} summarizes our measurements and Fig.~\ref{fig:delayplot} shows the relative precision on the time delays that can be achieved in one or two seasons of monitoring and how this compares with previously published delays. All light curves presented in this work are available on the online web application D3CS\footnote{\url{https://obswww.unige.ch/~millon/d3cs/COSMOGRAIL_public/}}, where they can be shifted in an interactive way to obtain an initial guess of the time delays. 

\begin{figure}[htbp!]
    \centering
    \includegraphics[width=0.49\textwidth]{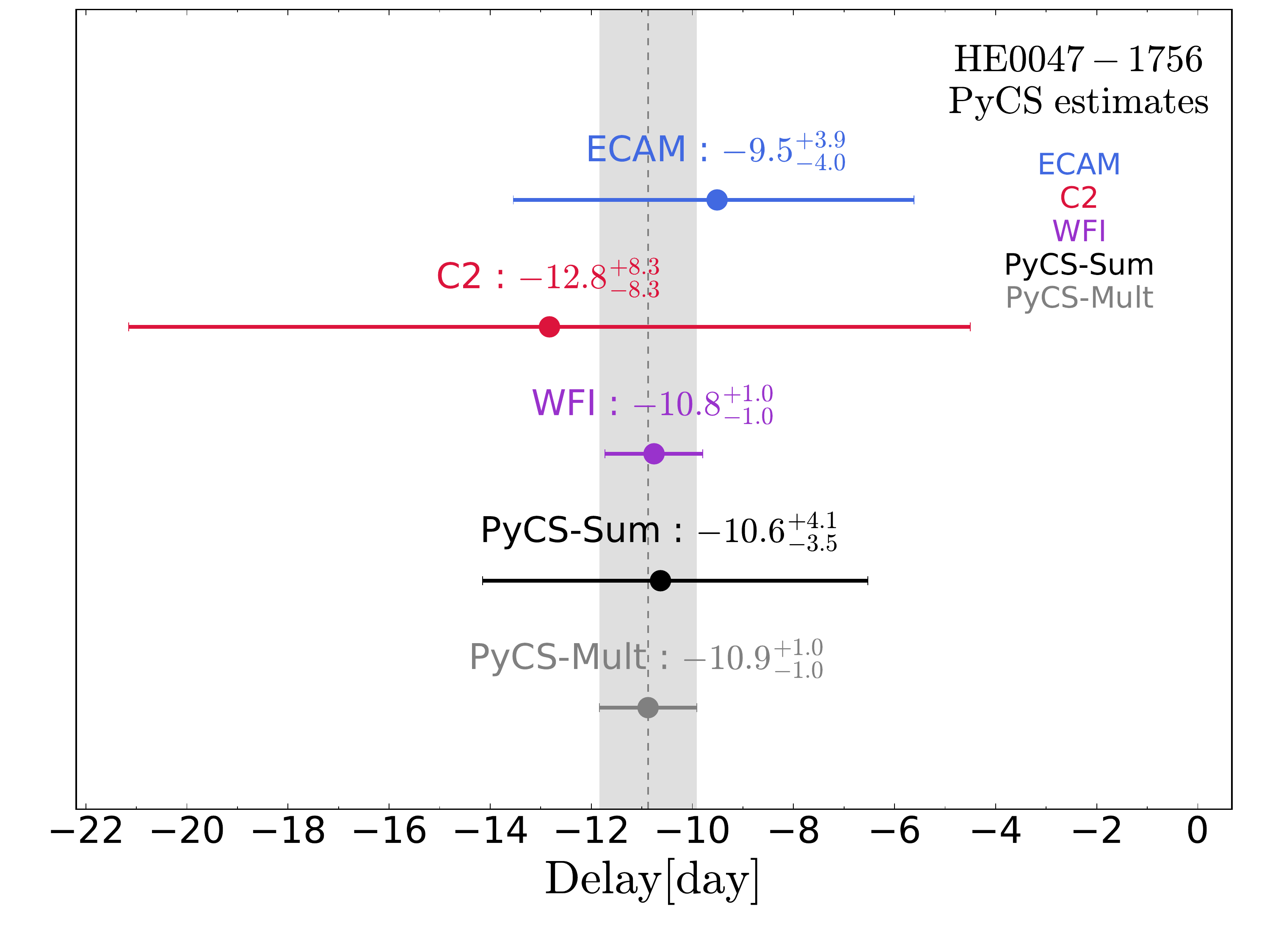}
    \caption{Time-delay estimate of \HEzerozero. Each point corresponds to the results of \pycs applied on a different data set. The ``PyCS-sum'' (in black) and the ``PyCS-mult'' (in shaded gray) are two possible combinations of the results of this work with the C2 and ECAM results measured in \cite{Millon2020}. ``PyCS-mult'' corresponds to the multiplication of the probability distribution, whereas ``PyCS-sum'' is their marginalization.\label{fig:delay_combi}}
    
\end{figure}

\subsection{\HEzerozero}
\HEzerozero was monitored during one and a half seasons. At least three very prominent features can be unambiguously detected in both the A and B light curves. Our final time-delay estimate is $\tau_{AB} = -10.8 ^{+1.0}_{-1.0}$ d (9.3\% precision), by combining the two \pycs estimators. This new measurement is within the 2$\sigma$ interval of a previous measurement by \cite{Giannini2017}, who found $\tau_{AB} = -7.6 \pm 1.8$ d with five seasons of monitoring at the 1.54 m Danish telescope at ESO La Silla observatory. The small discrepancy could be explained by the fact that the curve-shifting technique used in that work does not explicitly account for microlensing variation, which can possibly lead to underestimated uncertainties. The authors also report another estimate of the time delay measured with the free-knot spline technique of \pycs $\tau_{AB} = -7.2 \pm 3.8$, which accounts for extrinsic variation. This estimate yields larger uncertainties and is compatible within 1$\sigma$ with our measurement.

The time delay of \HEzerozero was also measured in \cite{Millon2020}, who found $\tau_{AB} = -10.4 ^{+3.5}_{-3.5}$ d using six seasons of monitoring with the C2 camera and eight seasons with the ECAM camera successively installed on the Euler telescope. As the same analysis framework was applied on these last two data sets and they do not cover the same period, we can consider these experiments to be independent and therefore combine the two time-delay estimates of \cite{Millon2020} with this new campaign conducted with the WFI instrument (see Fig.~\ref{fig:delay_combi}). We obtain in this way our final ``PyCS-mult'' estimate $\tau_{AB} = -10.9 ^{+0.9}_{-0.9}$ d (8.3\% precision). 

The precision of the measurement is significantly improved with high-cadence and high S/N data compared to the Euler monitoring campaigns, even though the duration of the monitoring is much shorter. The WFI images also have on average a better seeing than the ECAM and C2 data. This allows for a better deconvolution, especially for the B component, resulting in the B light curve being of much better quality than with the Euler telescope. Not surprisingly, this emphasizes the fact that the fainter component of each system dominates the final quality of the time-delay measurement.

\subsection{\WGzero}
The light curves of the quadruply imaged quasar \WGzero exhibit small-scale variations on the order of 0.05 - 0.1 mag visible in the three brightest images, A, B, and C. These small variations happen on timescales on the order of 20 to 40 days between MHJD = 58350 and MHJD = 58450, but are not visible in the D light curve because it is too noisy. However, two larger variations of the order of 0.2 mag are also visible in all four images at the end of the first season and during the second season. These last features allow us to measure the time delays relative to image D. 

The best relative precision is achieved for the BC delay, where $\tau_{BC} = -14.2 ^{+2.7}_{-2.5}$ d (18.3\% precision). The longest time delay is between image B and image D, where $\tau_{BD} = -21.6 ^{+4.2}_{-5.0}$ d (21.3\% precision). We can forecast how the these time-delay uncertainties transfer to the $H_0$ inference if the time-delay measurement remains the dominant source of errors compared to modeling and line of sight errors. This is likely to be the case here since the line of sight and modeling errors are typically on the order of ~5\% \cite[see][for the error budget of the H0LiCOW lenses]{Wong2019}. Assuming Gaussian probability distribution, we estimate that the relative uncertainty that directly propagates into the Hubble constant is on the order of $\sim$13.0\% by combining the three time delays relative to image B independently\footnote{The residual covariance between the three independent time delays is expected to be small and is therefore ignored in the computation of the total time-delay error propagating to $H_0$. We only aim to provide a rough estimate of the constraining power of each system on $H_0$.}.

In spite of the very good agreement between the two \pycs estimators, the free-knot spline estimator yields significantly larger uncertainties than the regression difference. This might be because the free-knot spline estimator is more sensitive to the photometric noise than the regression difference, but the latter requires more inflection points in the light curves to obtain precise time-delay estimates. \WGzero has relatively noisy light curves compared to other objects, but the quasar is highly variable; this might explain the good performance of the regression difference. Other objects with high photometric precision, but only a few inflection points in the light curves, such as \PSseize and \deuxM, exhibit the opposite behavior, that is, the free-knot spline estimator yields the best precision.  

%discuss why WG0214 is not that good
\WGzero was monitored for two seasons because the first season alone was not sufficient to measure any time delays at a precision better than 30\% despite intrinsic variations being clearly visible. The need of a second season for this object is explained by i) \WGzero is relatively faint so the photometric precision achieved in 30 min of exposure is lower than for other brighter objects (see Table \ref{tab:noise} for description of the photometric noise); ii) \WGzero is a compact quad (largest image separation is $1\arcsec85$), which makes it more sensitive to deconvolution errors, again increasing the photometric noise in the light curves; and iii) \WGzero has short time delays making it harder to obtain a good relative precision measurements. 
A third season of monitoring might be necessary to improve the time-delay precision and to make this system more valuable for time-delay cosmography. Still, this would be three times faster than with the previous COSMOGRAIL cadence and S/N on a 1 m-class telescope. Ideally, we aim for a precision below 5\% on the time-delay measurement, which is the threshold where the time-delay error becomes subdominant compared to the modeling and line-of-sight errors \citep[][]{Suyu2014, Suyu2017, Wong2019}.

%\todo{ALL}{Check the above two paragraphs}

\subsection{\DESzero}
Only one feature is visible in the B light curve of \DESzero around MHJD = 58500 d. This feature can be matched with the drop in the A light curve around MHJD = 58370 d.
Using \pycs, we obtained a final time-delay estimate of $\tau_{BD} = -128.4 ^{+3.5}_{-3.8}$ d (2.8\% precision). The long time delay of this system allowed us to reach a good relative precision although the overlap between the curves is limited. This object already has a sufficiently precise time-delay measurement to use it for time-delay cosmography. Although doubly imaged quasars are less effective, in principle, in constraining lens models, deep high-resolution images may reveal prominent and constraining rings due to the lensed host galaxy of the quasar as in \citet{Birrer2019}. 

\subsection{\deuxM}
\deuxM is a very bright quadruply imaged quasar discovered by \citet{Lucey2018}. The monitoring started shortly after the announcement of the discovery. Very small variations on the order of $\sim$40 mmag (peak-to-peak) are clearly visible in all light curves. The S/N in the light curves is sufficient to record even smaller variations on the order of $\sim$10 mmag in the three brightest multiple images A, B, and C. 

%paragraph on time delays: 
The most precise time delay is the B-C delay where $\tau_{BC} = +38.9 ^{+2.2}_{-2.2}$ d (5.7\% precision). We also measured at least one time delay relative to image A and image B with a precision better than 10\%, $\tau_{AB} = -30.5 ^{+2.2}_{-2.3}$ d (7.4\% precision) and $\tau_{CD} = -80.5 ^{+6.2}_{-8.4}$ d (9.1\% precision). Combining the three independent delays relative to image B and assuming Gaussian probability distribution, the total time-delay error that propagates to the Hubble constant is $\sim$4.4\%, making this object a promising target for future time-delay cosmography analysis\footnote{The time-delay error might not the dominant source of errors at this level of precision so the Gaussian approximation might not be sufficient for this object. Therefore, the total time-delay error given in this work is only an approximation.}. However, the lens redshift in \deuxM is yet unknown and might be difficult to measure from the ground owing to the high contrast between the bright quasar images and the faint lens galaxy.  

\subsection{\PSseize}
\PSseize does not show fast varying features, even in the A light curve, which has the best S/N. However, slow variations over the monitoring season allow us to obtain time-delay estimates with a precision below 30\% for the three brightest images. We measured 
$\tau_{AB} = -10.4 ^{+2.3}_{-2.2}$ d, $\tau_{AC} = -29.2 ^{+4.4}_{-5.1}$ d, and  $\tau_{BC} = -19.3 ^{+4.2}_{-4.9}$, that is, a 21.6\%, 16.3\%, and 23.6\% precision, respectively. We also measured $\tau_{AD} = -45.7 ^{+11.1}_{-10.7}$ d, but this time delay relative to image D is uncertain as a result of the lack of fast variation that can unambiguously be matched in all light curves. The fact that we rely on the slow variation of the quasar to measure the time delay and that the D light curve is relatively noisy makes the time-delay estimates relative to image D more dependent on the choice of estimator parameters and on the flexibility of microlensing model. As a consequence of this degeneracy between the slow intrinsic variation of the quasar and the slow microlensing variation, the time-delay probability distribution is multimodal, with a second peak appearing around -60 days. This second possibility however is less likely.

The combined time-delay error obtained by multiplying the two secure and independent delays relative to image A and using a Gaussian approximation is $\sim$13.0\%. This corresponds to the error that directly propagate to $H_0$ if the time-delay error remains the dominant source of uncertainties. These constraints are not yet sufficient for a competitive measurement of the Hubble constant with this system, but a second season of monitoring is likely to improve the precision given the continuous variations seen in the quasar. This will also help us better disentangle the microlensing and intrinsic variation in image D and allow us to discriminate between the two possible solutions for time delays relative to image D. The lens redshift is also unknown for this object, but the contrast between the lens and the quasar images is much lower than in \deuxM, so that a redshift determination should be easier.

\subsection{\DESvingt}
\DESvingt presents a quasar variation with a rise of 0.2 mag in image A in only $\sim$70 d between MHJD=58270 d and MHJD=58340 d. This feature is also clearly seen in the B light curves and allows us to measure $\tau_{AB} = +43.8 ^{+4.5}_{-4.0}$ d (9.7\% precision). We note a slight tension between the regression difference and the free-knot spline estimator at a statistical significance level of 1.1$\sigma$. In the residual $A-B$ curve, a slowly decreasing trend is visible at the beginning of the monitoring season, which might be attributed to microlensing in one of the two multiple images. As the regression difference estimator does not explicitly account for extrinsic variation whereas the free knot-spline estimator does, the small discrepancy between the two estimators could be explained by the presence of slow microlensing in the light curve. 
%We note that the measured time delay is in good agreement with the predicted delay from the lens model of \cite{Ostrovski2017} who found $\tau_{AB} = +52 \pm 11$ days, using a single isothermal ellipsoid in flat-\lcdm cosmology with $\Omega_m = 0.3$ and $H_0 = 70.0$\ksmpc.

%Added by TeX Support
\begin{table*}[htbp]
\caption{\label{tab:tdelays}Measured time delays, in days, for the two \pycs estimators and their combination (see text). In the case of \HEzerozero, the final \pycs-mult estimate is $\tau_{AB} = -10.9 ^{+0.9}_{-0.9}$d and is obtained by combining our WFI data set with monitoring data from the Leonhard Euler 1.2 m Swiss telescope \citep{Millon2020}.}
\centering
\renewcommand{\arraystretch}{1.3}
\begin{tabular}{c|c|c|c}
            & \pycs free-knot splines               & \pycs regression differences          & \pycs combined                        \\ \hline \hline
\HEzerozero & $\tau_{AB} = -10.7 ^{+1.2}_{-0.9}$    & $\tau_{AB} = -10.8 ^{+0.9}_{-0.9}$    & $\tau_{AB} = -10.8 ^{+1.0}_{-1.0}$         \\

\WGzero     & $\tau_{AB} = 7.3 ^{+4.6}_{-4.6}$     & $\tau_{AB} = 7.6 ^{+1.7}_{-1.7}$     & $\tau_{AB} = 7.5 ^{+2.7}_{-2.9}$             \\
            & $\tau_{AC} = -6.7 ^{+6.2}_{-6.2}$     & $\tau_{AC} = -6.6^{+2.0}_{-2.0}$      & $\tau_{AC} = -6.7^{+3.6}_{-3.6}$          \\
            & $\tau_{AD} = -15.7 ^{+6.9}_{-6.9}$  & $\tau_{AD} = -13.3 ^{+3.3}_{-3.3}$    & $\tau_{AD} = -14.1 ^{+4.4}_{-5.4}$       \\
            & $\tau_{BC} = -14.0 ^{+3.9}_{-3.9}$  & $\tau_{BC} = -14.3 ^{+1.8}_{-1.6}$    & $\tau_{BC} = -14.2 ^{+2.7}_{-2.5}$       \\
            & $\tau_{BD} = -23.0 ^{+6.5}_{-6.5}$  & $\tau_{BD} = -21.0 ^{+3.1}_{-3.1}$    & $\tau_{BD} = -21.6 ^{+4.2}_{-5.0}$       \\
            & $\tau_{CD} = -9.0 ^{+6.6}_{-6.6}$  & $\tau_{CD} = -6.8 ^{+3.1}_{-3.1}$    & $\tau_{CD} = -7.5 ^{+4.3}_{-5.1}$       \\
              
\DESzero    & $\tau_{AB} = -129.4 ^{+4.0}_{-4.5}$   & $\tau_{AB} = -127.8^{+3.0}_{-3.0}$    & $\tau_{AB} = -128.4^{+3.5}_{-3.8}$         \\

\deuxM      & $\tau_{AB} = -30.3 ^{+1.8}_{-2.0}$    & $\tau_{AB} = -30.7 ^{+2.6}_{-2.6}$    & $\tau_{AB} = -30.5^{+2.2}_{-2.3}$        \\
            & $\tau_{AC} = 8.7 ^{+1.1}_{-1.1}$      & $\tau_{AC} = 8.3 ^{+1.8}_{-2.0}$      & $\tau_{AC} = 8.6 ^{+1.4}_{-1.5}$           \\
            & $\tau_{AD} = -69.5 ^{+3.8}_{-5.2}$    & $\tau_{AD} = -76.3 ^{+8.7}_{-8.2}$    & $\tau_{AD} = -71.9 ^{+5.9}_{-8.5}$       \\
            & $\tau_{BC} = 39.0 ^{+1.7}_{-1.7}$  & $\tau_{BC} = 38.9 ^{+2.7}_{-2.9}$    & $\tau_{BC} = 38.9 ^{+2.2}_{-2.2}$      \\
            & $\tau_{BD} = -38.9 ^{+4.3}_{-6.7}$  & $\tau_{BD} = -45.6 ^{+9.0}_{-8.8}$    & $\tau_{BD} = -41.5 ^{+7.0}_{-8.8}$       \\
            & $\tau_{CD} = -78.2 ^{+4.2}_{-5.4}$  & $\tau_{CD} = -84.6 ^{+8.8}_{-8.2}$    & $\tau_{CD} = -80.5 ^{+6.2}_{-8.4}$        \\
            
\PSseize    & $\tau_{AB} = -10.6 ^{+1.7}_{-1.9}$    & $\tau_{AB} = -10.0 ^{+3.2}_{-2.1}$    & $\tau_{AB} = -10.4 ^{+2.3}_{-2.2}$          \\
            & $\tau_{AC} = -28.2 ^{+4.7}_{-2.6}$    & $\tau_{AC} = -30.7^{+5.6}_{-6.0}$     & $\tau_{AD} = -29.2^{+4.4}_{-5.1}$          \\
 %           & $\tau_{AD} = -41.8 ^{+5.3}_{-5.3}$    & $\tau_{AD} = -49.2 ^{+13.1}_{-12.2}$  & $\tau_{AD} = -44.0 ^{+7.4}_{-11.4}$      \\
            & $\tau_{BC} = -17.6 ^{+4.4}_{-2.6}$  & $\tau_{BC} = -21.5^{+4.9}_{-4.6}$    & $\tau_{BC} = -19.3 ^{+4.2}_{-4.9}$       \\
  %          & $\tau_{BD} = -31.4 ^{+5.5}_{-5.5}$  & $\tau_{BD} = -39.4 ^{+13.1}_{-12.6}$    & $\tau_{BD} = -33.8 ^{+7.6}_{-11.8}$      \\
  %          & $\tau_{CD} = -12.5 ^{+6.2}_{-6.2}$  & $\tau_{CD} = -16.0 ^{+10.8}_{-11.2}$    & $\tau_{CD} = -13.8 ^{+7.9}_{-8.9}$     \\
            
\DESvingt  & $\tau_{AB} = +41.3 ^{+3.3}_{-2.5}$ & $\tau_{AB} = +46.6 ^{+3.7}_{-3.2}$ &                      $\tau_{AB} = +43.8 ^{+4.5}_{-4.0}$        \\
\end{tabular}
\end{table*}

\section{Residual fast extrinsic variability}

By shifting the light curves by their measured time delays and subtracting them pair-wise, we obtained difference light curves, which highlight the residual extrinsic variations. During this process, we did not correct for any microlensing variability. We observe in the $B-A$ difference light curve of \WGzero a fast variation on the order of 0.1 mag around MHJD = 58480 and happening on a timescale of only 20 days. We observe a similar effect in \deuxM, where small variations on the order of 10 mmag in the $B-A$  difference light curve are visible at the beginning of the monitoring season. 

Although these variations could be a signature of fast microlensing, the fact that this happens at the same time as an intrinsic variation that is visible in all multiple images might also indicate that an additive flux component is contaminating one or both images. To verify that an additive flux component does not impact the measured time delays, we fit an additional parameter corresponding to a constant shift in flux of the light curves. In practice, this corresponds to a stretch in magnitude, i.e. along the y-axis in Fig.~\ref{fig:lcs}. This flux shift differs from a shift in magnitude that we normally apply to the light curves and that corresponds to the multiplicative (flux) factor produced by the lensing magnification.

We applied this flux correction to \deuxM and \WGzero, which are the two objects the most affected by this effect. This reduces the amplitude of the variations seen in the difference curves but does not remove them completely. Still, we applied our time-delay measurement pipeline to the corrected data. This only changes the measured time delay marginally and none of the measured time delays are shifted by more than the reported uncertainties. The maximal changes over the six measured time delays for each object corresponds to 0.4$\sigma$ for \WGzero and 0.7$\sigma$ for \deuxM. We thus conclude that the distortions of the light curves that we observe in these two lens systems do not significantly impact the measured time delays.

Although instrumental effects or residual contamination after the deconvolution could be a possible explanation for the observed distortion of the light curves, this might also come from the regions of multiple source sizes contributing to the $R$-band flux and being differently microlensed. Indeed, each lensed image is composed of a variable component (central accretion disk) and a nonvariable component; that is, the broad line region (BLR) and the central part of the bulge of the host galaxy. The latter is little or not affected at all by microlensing because its size is much larger than microcaustics. Thus, if microlensing affects the variable part of one image but not the other, this would produce variations of larger amplitude in the microlensed image and hence result in residuals in the difference light curve. A description of a similar ``differential amplification'' effect can be found in Sect.~3.3.3 of \cite{Sluse2006}. The lens light could also contribute to the nonmicrolensed component that is needed to produce the effect. Finally, we note that the nonmicrolensed component might also be variable as a result of the reverberation of the continuum emission in the BLR as suggested by \cite{Sluse2014}.

%If i) the inner part accretion disk is microlensed in image B but not in the other images and, if ii) the total flux is the sum of two components, for example, a variable compact component (e.g, the accretion disk) and a nonvariable extended component, which is not affected by microlensing because of its large source size (e.g., the Broad Line Region (BLR) or the bulge of the host galaxy), the intrinsic variation of the quasar will then look sharper in image B because the variable component is magnified by the microlenses. Therefore, the ratio between the variable and stable component is larger than in the other multiple images, leading to larger amplitude of the intrinsic variation in image B. A description of a similar ``differential amplification" effect can be found in Sect 3.3.3 of \cite{Sluse2006}. We note that the lens light could also play the role of nonmicrolensed component that is needed to produce the effect. The nonmicrolensed component might also be variable as a result of the reverberation of the continuum emission in the BLR as suggested by \cite{Sluse2014}. 
%\todo{ALL}{Any ideas from relevant work to be cited ? Tie and Kochanek maybe but that might be another level of un-needed complexification of the paper...}

Our new high-quality light curves probably point to new subtle differential microlensing effects that were unseen with data of lower quality. In the present paper, we limit ourselves to checking whether these effects impact time-delay cosmography, and we show that they do not. However, our data may allow us to study quasar structure on very small physical scales and at cosmological distances. This is beyond the scope of this paper but we point to a potential opportunity to use high-cadence and high S/N multiband light curves to scrutinize the inner regions of quasars and their host galaxies with microlensing.

%However, it will be very interesting to use these new subtle effects in the future to studWhile these studies are beyond the scope 

%Although a further study of the origin of this distortion might be of importance to further understand the emission mechanism of Active Galactic Nuclei (AGN), this is beyond the scope of this paper. Source size effects that might be detectable here thanks to the very high quality of the monitoring data, could be a unique opportunity to study the structure of AGN on very small physical scale and at cosmological distances. However, multi-band monitoring data at high cadence will be necessary to further understand this effect. 

%====================
\section{Conclusions}
%====================
\label{section:conclusion}

We present the results of the first intensive high-cadence and high S/N monitoring campaign in the framework of the TDCOSMO collaboration. We measured new time delays in three doubly imaged and three quadruply imaged quasars using data taken almost daily  with the MPIA 2.2 m telescope at ESO observatory, La Silla. The most precise delay is obtained for \DESzero, where $\tau_{AB} = -128.4 ^{+3.5}_{-3.8}$ d (2.8\% precision). All other objects have at least one time delay measured with a precision better than 18.3\%, including systematics due to the residual extrinsic variability. \PSseize presents the most uncertain estimates owing to the absence of fast intrinsic variation. For this object, a second season of monitoring will be necessary in order to reach uncertainties on the order of $\sim$10 \% on the best measured time delay.

We confirm that high-cadence and high S/N monitoring data with 2 m-class telescopes can provide precise time delay in one single season, as was first explored by \citet{Courbin2017}. This observation strategy allows us to better disentangle microlensing from the intrinsic signal of the quasar by recording its small-amplitude and fast variations. The unprecedented quality of the data also allows us to detect small distortions of the light curves between the multiple images, which are not only shifted in time and in magnitude but also stretched along the magnitude axis. This effect is detected in two lensed systems, namely \deuxM and \WGzero. We suggest that a source size effect might explain this distortion if the broadband emission contains flux arising from the compact active galactic nucleus continuum and from a spatially more extended region, such as the BLR or the bulge of the host galaxy. The differential microlensing between those two sources of emission may explain the observed signal. Although the exact origin of this effect remains to be clarified, we can still correct for the contaminating component and find that this does not change the measured time delays. 
%\todo{Fred}{Check the above paragraph}

We used two time-delay estimators in the \pycs package, namely the regression difference and the free-knot spline. We note a very good agreement between these two estimators overall, which indicates that the choice in the modeling of the extrinsic variability does not significantly impact the final time-delay estimates. When available, we also include monitoring data from the Leonhard Euler 1.2 m Swiss telescope from \cite{Millon2020}. We combined the measurements to obtain the time delay of \HEzerozero, $\tau_{AB} = -10.9 ^{+0.9}_{-0.9}$ d with 8.3\% precision. 

As the number of known lensed quasar is increasing quickly with new wide-field surveys, the rapid follow-up of the newly discovered quasars is crucial to turn the corresponding new time delays into cosmological constraints. The Rubin Observatory Legacy Survey of Space and Time (LSST) will provide high S/N monitoring data for a large part of the sky but its cadence will be limited to one point every few days in any given band. Our observations emphasize that the highest possible temporal sampling is just as important as S/N to overcome the microlensing variability. It is therefore likely that LSST light curves will require complementary data from 2 m-class telescopes or larger with a daily cadence.

\begin{acknowledgements}
 We warmly thank R. Gredel, H.-W. Rix and T. Henning for allowing us to observe with the MPIA 2.2m telescope and for ensuring the daily cadence of the observations is achieved. This program is supported by the Swiss National Science Foundation (SNSF) and by the European Research Council (ERC) under the European Union’s Horizon 2020 research and innovation program (COSMICLENS: grant agreement No 787886). TT and CDF acknowledge support by NSF through grant “Collaborative Research: Toward a 1\% Measurement of The Hubble Constant with Gravitational Time Delays  AST-1906976. CDF acknowledges support for this work from the National Science Foundation under Grant No. AST-1907396. SHS and DCYC thank the Max Planck Society for support through the Max Planck Research Group for SHS. T.A. acnkowledges support from Proyecto Fondecyt N 1190335 and the Ministry for the Economy, Development, and Tourism's Programa Inicativa Cient\'ifica Milenio through grant IC 12009. This research made use of Astropy, a community-developed core Python package for Astronomy \citep{Astropy2013, Astropy2018} and the 2D graphics environment Matplotlib \citep{Hunter2007}.

\end{acknowledgements}

\bibliographystyle{aa}
\bibliography{biblio}

\end{document}